%% file: ANA-EXOT-2019-33-PAPER.tex
\pdfoutput=1
\documentclass[PAPER,cernpreprint, atlasdraft=false, texlive=2023, UKenglish, texmf, orcidlogo]{atlasdoc}
 
\usepackage{atlaspackage}
\usepackage{atlasbiblatex}
 
\usepackage{atlasphysics}

\graphicspath{{logos/}{fig/ATLAS/}}

\AtlasTitle{Search for magnetic monopoles and stable particles with high electric charges in $\sqrt{s}=13$~TeV~$pp$ collisions with the ATLAS detector}

\AtlasAbstract{
We present a search for magnetic monopoles and high-electric-charge objects using LHC Run~2 $\sqrt{s}=13$~TeV proton--proton collisions recorded by the ATLAS detector. A total integrated luminosity of $138$~fb$^{-1}$ was collected by a specialized trigger. No highly ionizing particle candidate was observed. Considering the Drell-Yan and photon-fusion pair production mechanisms as benchmark models, cross-section upper limits are presented for spin-0 and spin-\textonehalf\ magnetic monopoles of magnetic charge $1g_\textrm{D}$ and $2g_\textrm{D}$ and for high-electric-charge objects of electric charge $20 \leq |z| \leq 100$, for masses between 200~GeV and 4000~GeV. The search improves by approximately a factor of three the previous cross-section limits on the Drell-Yan production of magnetic monopoles and high-electric charge objects.
Also, the first ATLAS limits on the photon-fusion pair production mechanism of magnetic monopoles and high-electric-charge objects are obtained.
}

\AtlasRefCode{EXOT-2019-33}
 
\PreprintIdNumber{CERN-EP-2023-150}

\arXivId{2308.04835}
 
\HepDataRecord{10.17182/hepdata.141286}
 
\AtlasJournalRef{J. High Energ. Phys. 2023, 112 (2023)}
\AtlasDOI{10.1007/JHEP11(2023)112}

\hypersetup{pdftitle={ATLAS document},pdfauthor={The ATLAS Collaboration}}
 
\begin{document}
 
\maketitle

\section{Introduction}
 
The magnetic monopole, an isolated magnetic charge,
is appealing in that it restores the broken electric--magnetic dual symmetry in Maxwell's equations. Furthermore, Dirac's theory of magnetic monopoles~\cite{Dirac1931,Dirac1948}, which is compatible with quantum mechanics,
offers an explanation for the quantization of electric charge. A
Dirac magnetic monopole is a fundamental point particle that may take any spin or mass. 
In contrast, the monopoles predicted by Grand Unification Theories (GUTs)~\cite{Polyakov1974,tHooft1974}, typically produced cosmologically via the Kibble mechanism~\cite{Kibble1976}, are extended objects with large mass $\sim10^{17}$~GeV.
Recent developments in extensions of the Standard
Model~\cite{Kephart2017,CHO1997360,Kimm_2015,Ellis2016,Arunasalam2017,Ellis2017,Arai2018,Mavromatos2017,Mavromatos2018,Cho2019,Hung2020,Ellis2020} predict "electroweak" monopoles, which appear as gauge bosons of particle interactions before spontaneous symmetry breaking, and thus
can have a mass as low as the TeV scale~\cite{Kimm_2015,Arunasalam2017,Ellis2017,Arai2018}.
 
Dirac's quantization condition predicts that the magnetic charge $g$ of a monopole is an integer multiple of
the fundamental magnetic charge $g_\textrm{D}$, which is given by
\begin{equation}
g_\textrm{D} =\frac{e}{2\alpha}\approx 68.5e,
\end{equation}
in natural SI units, where  $\alpha \approx 1/137$ is the fine structure constant and $e$ is the unsigned electron charge. 
This implies that the energy loss, or stopping power, of a high-velocity Dirac monopole of magnetic charge $|g|=g_\textrm{D}$ in matter is similar to that of an ion with electric charge $|z| = 68.5$, where $z$ is in units of $e$.
Hence, magnetic monopoles, along with high-electric-charge objects (HECOs), such as strange and up-down quark matter~\cite{Chin1979,Farhi1984,udQM2018}, and Q-balls~\cite{Coleman1985,Kusenko1997},
are collectively known as highly ionizing particles (HIPs). A consequence of the high ionization is the production of a large number of $\delta$-rays, which are energetic electrons emitted along the HIP trajectory. The characteristic signature of HIPs, therefore, is high ionization coupled with a large number of $\delta$-rays.

Searches for magnetic monopoles and HECOs have been extensively performed in cosmic rays~\cite{Battistoni:1983,Krishnaswamy:1984,Orito:1991,Soudan-2:1992,MACRO:1994,MACRO:1997,MACRO:2000,MACRO:2002a,Baksan:2006,BAIKAL:2008,SLIM:2008a,SLIM:2008b,RICE:2008,AMANDA:2010,ANITA-II:2010,PierreAuger:2016,ANTARES:2012,ANTARES:2017,ANTARES:2021b,ANTARES:2021a,ANTARES:2022,ANTARES:2023,IceCube:2013,IceCube:2016a,IceCube:2016b,IceCube:2018,IceCube:2022,NOvA:2015,NOvA:2021} and at colliders~\cite{PETRA:1983,PEP:1982,PEP:1984,CLEO:1987,TASSO:1988,TRISTAN:1988,TRISTAN:1989,OPAL:2008,MODAL:1992,L6-MODAL:1993,H1:2004,Johnsen:1973fop,CDF:2005,Kalbfleisch:2003,Kalbfleisch:2000,D0:1998,Price:1987,Price:1990,Bertani:1990,Hoffmann:1978,Carrigan:1977,Aubert:1982,Carrigan:1973,Carrigan:1974,PhysRevLett.79.3134,STAR:2005,E886:1996,E864:1999,Xu_E864:1999,EXOT-2011-27,EXOT-2014-16,EXOT-2017-20}, as reviewed in Refs.~\cite{Giacomelli2003,Milton2006,Burdin2015,Patrizii2015,Patrizii2019,Mavromatos2020}.
Direct cosmic-ray searches look for signals of relic GUT monopoles produced in the early universe upon symmetry breaking~\cite{Kibble1976} and nuclearites (nuggets of strange quark matter in collision with the Earth) that were either formed in the early universe or during collisions of neutron stars~\cite{DeRujula1984}. Limited by the collision energy, collider searches for HIPs target stable Dirac magnetic monopoles, Q-balls and quark matter with masses in the TeV range. Since the searches for monopoles of cosmological origin have limited sensitivity to TeV-mass HIPs, the cross-section limits for low-mass HIPs from searches at colliders are 6–9 orders of magnitude more stringent.
 
The first Large Hadron Collider (LHC) search for monopoles,
carried out by the ATLAS Collaboration with \SI{7}{\tera\ev} proton-proton ($pp$) collision data in Run~1~\cite{EXOT-2011-27}, set a precedent for later LHC searches for HIPs~\cite{EXOT-2014-16,EXOT-2017-20,EXOT-2012-10,EXOT-2013-20,EXOT-2017-13,	ATLAS2020a,CMS-EXO-12-026,CMS-EXO-15-010, MoEDAL2016a,MoEDAL2017a,MoEDAL2018,MoEDAL2019,MoEDAL2021Schwinger,dyonMoEDAL2021,MoEDAL2021HIP}.
The ATLAS monopole searches are complementary to those performed using the dedicated MoEDAL experiment at the LHC. While MoEDAL is able to probe magnetic charges up to $5g_\textrm{D}$~\cite{MoEDAL2016a,MoEDAL2017a,MoEDAL2018,MoEDAL2019,MoEDAL2021Schwinger}, its acceptance is limited by its forward location at $2<\eta<5$. Consequently, the ATLAS search, which considers $|\eta|<1.375$, has significantly higher sensitivity for $1g_\textrm{D}$ and $2g_\textrm{D}$, the charge range in which it has a good acceptance.
While ATLAS is sensitive to HECOs with electric charges up to $|z| =100$, MoEDAL searches for HECOs with electric charges up to $|z| = 175$~\cite{MoEDAL2021HIP}. MoEDAL has also searched for dyons (particles with both electric and magnetic charges) with magnetic charges up to 5$g_\textrm{D}$ and electric charges up to $|z| = 200$~\cite{dyonMoEDAL2021}.
 
The ATLAS experiment has considered the highly ionizing signature
to probe magnetic monopoles in the range $0.5g_\textrm{D}<|g|<2g_\textrm{D}$ and HECOs in the range $20\le |z| \le 100$ in previous searches~\cite{EXOT-2011-27,EXOT-2014-16,EXOT-2017-20}, the most recent of which used \SI{34.4}{\per\fb} of $\sqrt{s}=$~\SI{13}{\tera\ev} $pp$ collision data~\cite{EXOT-2017-20}.
The present HIP search uses data from $\sqrt{s}=$~\SI{13}{\tera\ev} LHC proton-proton collisions recorded by the ATLAS detector
between 2015 and 2018 during Run 2. A custom HIP trigger
was implemented in October~2015, and collected a total integrated luminosity of \SI{138}{\per\fb}.
The results presented
supersede those of Ref.~\cite{EXOT-2017-20}, benefiting from a four-fold increase in data statistics
and the addition of $Z^0$ exchange in the spin-\textonehalf\ Drell-Yan HECO production model~\cite{Song2022}.
New for this search are results for HIP production by the photon-fusion  mechanism~\cite{Kurochkin2006,Dougall2009,Epele2012,Baines2018,Song2022}, which has a higher predicted cross-section than that of Drell-Yan production in $\sqrt{s}=$~\SI{13}{\tera\ev} $pp$ collisions, whereas the cross sections are comparable at $\sqrt{s}=$~\SI{8}{\tera\ev}.
 
The ATLAS detector is briefly covered in Section~\ref{sec:ATLAS}, the production models and simulation are described in Section~\ref{sec:MC}, the signal selection is discussed in Section~\ref{sec:signal}, the method to estimate the background is presented in Section~\ref{sec:bkg}, and the procedure to evaluate the systematic uncertainties is outlined in Section~\ref{sec:uncert}. Finally, the results are presented in Section~\ref{sec:results}, followed by the conclusions in Section~\ref{sec:conclusions}.
 
\section{The ATLAS Detector}
\label{sec:ATLAS}
 
The ATLAS detector~\cite{PERF-2007-01} is a multipurpose particle detector with a forward-backward symmetric cylindrical
geometry and a near $4\pi$ coverage in solid angle.\footnote{ATLAS uses a right-handed coordinate system with its origin at the nominal interaction point (IP) in the center of the detector and the $z$ axis coinciding with the axis of the beam pipe. The $x$ axis points from the IP to the center of the LHC ring, and the $y$ axis points upward. Cylindrical coordinates $(r, \phi)$ are used in the transverse plane, $\phi$ being the azimuthal angle around the beam pipe. The pseudorapidity is defined in terms of the polar angle $\theta$ as $\eta =-\ln \tan(\theta/2)$.} It comprises an inner tracking detector (ID) surrounded by a thin superconducting solenoid providing a 2~T axial magnetic field, electromagnetic
and hadron calorimeters, and a muon spectrometer.
To detect HIPs, it suffices to only consider the signals in the Transition Radiation Tracker (TRT)~\cite{IDET-2015-01} and the electromagnetic (EM) calorimeter.
The TRT has two sections: a barrel of radius $0.563~\text{m} <r<1.066~\text{m}$ covering the pseudorapidity range $0 < |\eta| < 1.06$ and end-caps at either end of the barrel covering $0.77 < |\eta| < 2.0$. In the barrel the 4-mm diameter TRT straws are aligned parallel to the beam pipe, while in the end-caps they are in a radial array. The straws are filled with a Xenon- or Argon-gas-based mixture.
An electron traversing the TRT induces transition radiation, which augments the energy recorded in a straw.
The energy deposited in a straw by ionizing particles or transition radiation is compared with two energy thresholds, \SI{200}{\ev} and \SI{6}{\kilo\ev} (\SI{200}{\ev} and \SI{2}{\kilo\ev} in Ar), to identify low threshold (LT) and high threshold (HT) hits, respectively. HT hits can indicate the passage of electrons or HIPs.
Argon absorbs the transition radiation with a reduced efficiency relative to that of Xenon. Consequently, the HT hit probability for electrons is 50\% lower in Ar than in Xe. However, the energy deposited by a HIP typically exceeds the high threshold in either gas.
 
The electromagnetic calorimeter, covering the $|\eta| < 1.475$ region, is composed of accordion-shaped lead absorbers and copper electrodes and filled with liquid Argon (LAr). The detector has four layers (labeled presampler, EM1, EM2 and EM3) of varying radiation depths at increasing radii from the beampipe, EM2 being the deepest (16$X_0$ to 20$X_0$). The smallest sensor unit, called a cell, has different dimensions in each layer, e.g., $\Delta\eta\times\Delta\phi=0.025\times0.025$ in EM2.

A two-level trigger system~\cite{TRIG-2016-01} is used to select events. The first-level trigger is implemented in hardware and uses a subset of the detector information to accept events at a rate below 100~kHz on average depending
on the data-taking conditions. This is followed by a software-based high-level trigger (HLT) that reduces
the rate of selected events to 1~kHz for offline storage. A custom HLT algorithm, explained in Section~\ref{sec:signal}, is used to recognize the highly ionizing signature of monopoles and HECOs. An extensive software suite~\cite{ATL-SOFT-PUB-2021-001} is used in data simulation, in the reconstruction and analysis of real and simulated data, in detector operations, and in the trigger and data acquisition systems of the experiment.
 
\section{Simulation}
\label{sec:MC}
 
\begin{figure}
\centering
\includegraphics[width=0.8\linewidth,keepaspectratio=true]{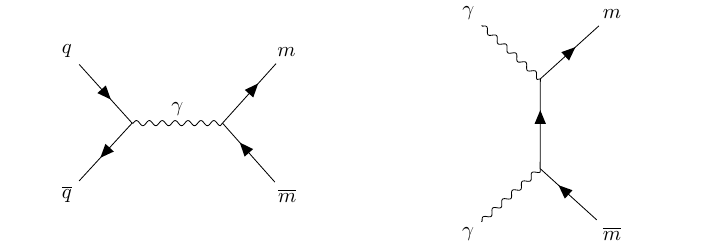}
\caption{Feynman diagrams for Drell-Yan (left) and photon-fusion (right) pair-produced monopoles $m$. HECO production occurs by analogous diagrams. Other production modes not shown here, but which can be seen in Ref.~\cite{Song2022}, include spin-0 HIP production by a four-point photon-fusion process and Drell-Yan production of spin-\textonehalf\ HECOs mediated by $Z^0$ exchange, in addition to photon exchange. $Z^0$ exchange is forbidden for spin-0 HECOs and for Dirac monopoles of either spin.}\label{fig:feynman}
\end{figure}

Drell-Yan (DY) and photon fusion (PF) are considered as benchmark models for pair production of spin-0 and spin-\textonehalf\ HIPs.
The model implementations are described in detail in Ref.~\cite{Song2022}.
As a result of the high HIP charge, the charge-squared dependence of the HIP coupling to the gauge boson leads to divergences in the perturbative expansion beyond leading order~\cite{Milton2006}. To provide an approximation to these highly non-perturbative processes, only leading-order interactions, as described by the Feynman-like diagrams seen in Figure~\ref{fig:feynman}, are considered. This approximation implies uncertainties in the pair-production cross section
predictions and in the HIP kinematic distributions. The former are only used to derive mass limits for comparison with other experiments. The latter dictate the selection efficiencies. Since the interaction of the HIPs with the detector is spin-independent, a comparison of the selection efficiencies for spin-0 vs spin-\textonehalf\ HIPs provides a measure of how model uncertainties affect the search acceptance.
 
Monte Carlo (MC) samples are produced for both production models for magnetic monopoles with magnetic charges $|g| = g_\textrm{D}$ and 2$g_\textrm{D}$ and HECOs with electric charges $|z|= 20$, 40, 60, 80 and 100. In all cases, the masses considered are 200, 500, 1000, 1500, 2000, 2500, 3000 and 4000~GeV. The available energy of the interaction between two quarks in the collision sets the upper mass constraint. Because HIPs with mass below \SI{200}{\giga\ev} are more probable in softer $pp$ collisions, in general they have lower kinetic energies and, consequently, poor selection efficiency.
The leading-order benchmark models are implemented in \textsc{MadGraph}5\_\textsc{aMC}@NLO 2.8.1~\cite{Alwall:2014hca}, which is used to generate the HIP pairs and to calculate the production cross sections.
The events were interfaced to \textsc{Pythia}~8.244~\cite{Sjostrand:2014zea,Sjostrand:2006za} to model the parton showering and hadronization, with parameter values set according to the A14 tune~\cite{ATL-PHYS-PUB-2014-021}, and the NNPDF2.3LO~\cite{NNPDF23LO} and LUX$_\textrm{QED}$~\cite{Manohar_2017} parton distribution functions were used for DY and PF production, respectively.

To determine selection efficiencies, samples of 60~000 MC events of Drell-Yan spin-\textonehalf\ HIPs of a given mass and charge were simulated using \textsc{Geant4}~\cite{Agostinelli:2002hh} and processed with the ATLAS reconstruction software. The \textsc{Geant4} simulation describes the HIP ionization energy loss in the ATLAS detector, the $\delta$-ray production and secondary ionization, and the monopole acceleration in the magnetic field. The standard Bethe-Bloch formula is used for ionization by HECOs, whereas a modified formula accounting for the velocity-dependent Lorentz force~\cite{Ahlen1978} is used for monopoles. In both cases, the energy loss by ionization is proportional to the square of the charge.
The simulation of highly ionizing particles in the ATLAS detector was validated by confirming the expected acceleration and trajectory in the magnetic field and the dependence of the ionization on the charge and on the speed of light fraction $\beta$. In addition, the simulation of the ionization and $\delta$-ray production of HIPs in liquid argon was validated against published ion beam data in liquid argon~\cite{HIPBirks}.
These MC samples also include “pileup”, which are additional simulated minimum bias collisions overlaid on each event according to the distribution of the number of $pp$ interactions per bunch crossing in the data.
 
Full simulation of HIP events is computationally intensive due to the high ionization and $\delta$-ray production. Single-particle events have less detector activity than events with pair production from proton-proton collisions, and consequently can be abundantly produced at a reasonable cost of computational resources. Hence, the selection efficiencies of DY spin-0 HIPs and PF pair-produced HIPs of both spins are extrapolated by sampling model-independent efficiency maps based on the HIP kinematic properties, transverse kinetic energy $E^\text{kin}_\text{T}=E_\text{kin}\sin\theta$ and pseudorapidity $|\eta|$. The efficiency maps of a given mass and charge are derived from samples of 120~000 events comprising single monopoles or HECOs simulated using \textsc{Geant4}, overlaid with pileup, and processed with the ATLAS reconstruction software. Given the statistics available in the single-particle samples, the bin size of the maps was optimized to ensure that most bins are populated.
 
\section{Signal Selection}
\label{sec:signal}
The selection of HIP-like events exploits their characteristic signature in the ATLAS detector.
When a HIP traverses a TRT straw, in addition to producing an HT hit, it generates many $\delta$-rays, which in turn result in additional HT hits in neighbouring straws.
In the mass and energy regime of this study, HIPs lose energy primarily via ionization, with radiation losses representing less than 5\% of the total energy lost~\cite{Ahlen1980}. Consequently, since HIPs do not induce a shower in the EM calorimeter, their EM energy deposits have low lateral dispersion.
The kinematic properties of the considered HIPs are such that most of them will stop before reaching the hadronic calorimeter. Consequently, the typical signature of a HIP in the ATLAS detector includes many TRT HT hits in a region aligned with a narrow high-energy deposit in the LAr EM calorimeter.
 
The HIP candidate selection begins with a custom HIP trigger.  First, the
standard first-level EM trigger requires an energy deposit of transverse energy $E_T>$~\SI{22}{\giga\ev} in the EM calorimeter. The EM trigger rejects hadrons by vetoing candidates with $E_T<$~\SI{50}{\giga\ev} in conjunction with an $E_T$-dependent minimum of \SI{1}{\giga\ev} to \SI{2}{\giga\ev} of energy in the hadronic calorimeter. 
The locations of any selected energy deposits determine regions of interest (ROI).
The custom HIP HLT determines the number, $N_\mathrm{HT,trig}$, of TRT HT hits and the fraction, $f_\mathrm{HT,trig}$, of TRT hits that are HT hits in a \SI{10}{\milli\radian} $r-\phi$ wedge around the first-level ROI position.
To control the rate while maintaining high signal efficiency for HIPs, the HIP HLT requires $N_\mathrm{HT,trig}> 30$ and $f_\mathrm{HT,trig}> 0.5$ in the wedge. In addition, it imposes a requirement on the pseudorapidity, $|\eta| < 1.7$, since the forward regions contain more multijet background events.
 
Events collected by the trigger are processed by the standard ATLAS reconstruction software, which performs a sophisticated topological clustering algorithm~\cite{PERF-2014-07} in the calorimeter. A cluster is a group of cells, neighbouring in 3D, in which significant energy deposition is recorded. EM clusters with $E_T>$~\SI{22}{\giga\ev} and $|\eta|<1.375$ are considered as a starting point for HIP signal selection. The $|\eta|<1.375$ requirement is needed to exclude the EM calorimeter end-caps, where the correlation between the final discriminating variables, described below, is enhanced.
 
\begin{figure}[htb]
\centering
\includegraphics[width=0.80\linewidth,keepaspectratio=true]{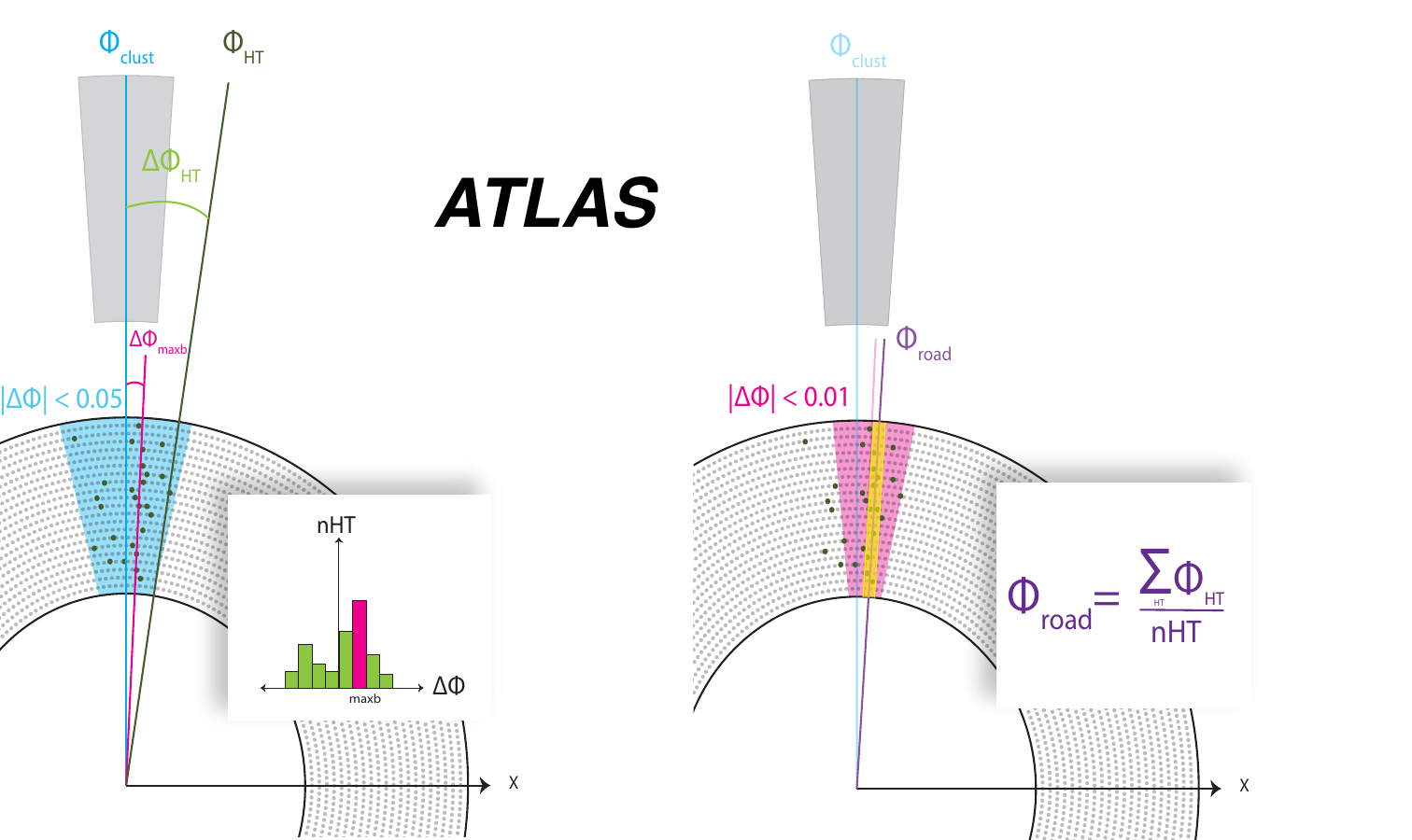}
\caption{Schematic depicting the TRT road-centering and hit-counting algorithms for HIP candidates. Centering of the road happens in two stages. In the first stage (left), a $\pm0.05$-rad
wedge (shaded blue) is defined around the azimuthal angle $\phi_\text{clust}$ of the EM cluster
(blue solid line). In that wedge, the HT hit $\phi_\text{HT}$ distribution with an 0.8-mrad bin size is determined. The center of the bin containing the largest number of HT hits defines the azimuthal angle $\phi_\text{maxb}$ (pink solid line). For the second stage (right), a new $\pm0.01$-rad wedge (shaded pink) is centered around $\phi_\text{maxb}$. The azimuthal angle $\phi_\text{road}$ (purple) is computed from the angular difference of each HT hit in this wedge with respect to $\phi_\text{clust}$. The final $\pm4$-mm road (shaded yellow) is defined around $\phi_\text{road}$.}\label{fig:hitcount}
\end{figure}

This analysis does not use ATLAS track reconstruction for two main reasons: first, the trajectories of magnetic monopoles moving in a solenoidal magnetic field bend in the $r-z$ plane (unlike charged particles, which bend in the $r-\phi$ plane), and second, the presence of multiple $\delta$-rays confuses the track fitting algorithm. Instead, HIPs are reconstructed in a manner similar to that of the HIP HLT: each EM cluster defines a starting $\phi$ position in the TRT for the hit-counting region; the $\phi$ position is fine-tuned to align with the region containing the highest number of HT hits; finally, an $8$-mm-wide rectangular road in the barrel is centered on that new $\phi$ position. This road width is sufficient to capture energy deposited by the HIP and any $\delta$-rays that may have penetrated a neighbouring TRT straw, but narrow enough to avoid counting nearby LT hits from pileup. In the end-cap, where the straws are radial, a \SI{12}{\milli\radian} $r-\phi$ wedge is used.
The numbers of TRT LT and HT hits, $N_\mathrm{LT}$ and $N_\mathrm{HT}$, respectively, in the road (wedge) aligned with the EM cluster are counted, as depicted in Figure~\ref{fig:hitcount}.
 
\begin{figure}[h!]
\centering
\includegraphics[width=0.80\linewidth,keepaspectratio=true]{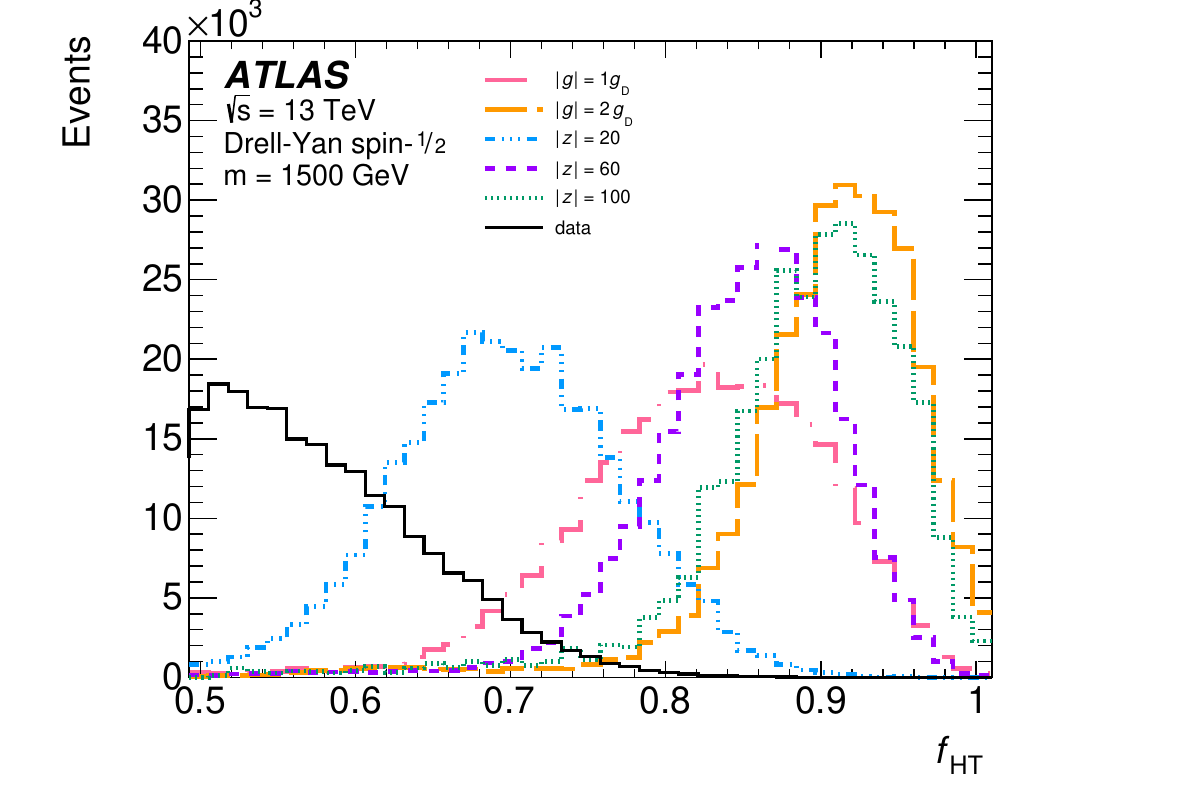}
\caption{Distribution of the $f_\textrm{HT}$ variable for events that passed the HIPTRT trigger and the criteria $|\eta| < 1.375$ and $N_\mathrm{HT}>30$. Only the TRT road candidate with the highest $f_\text{HT}$ value in each event is included in the plot. The black distribution corresponds to 20\% of the Run~2 data and the coloured distributions correspond to the mass 1500~GeV Drell-Yan spin-\textonehalf\ signal samples for various charges. All MC distributions have been normalized to the number of events in the data.}\label{fig:fht1d}
\end{figure}

\begin{figure}[h!]
\centering
\includegraphics[width=0.8\textwidth]{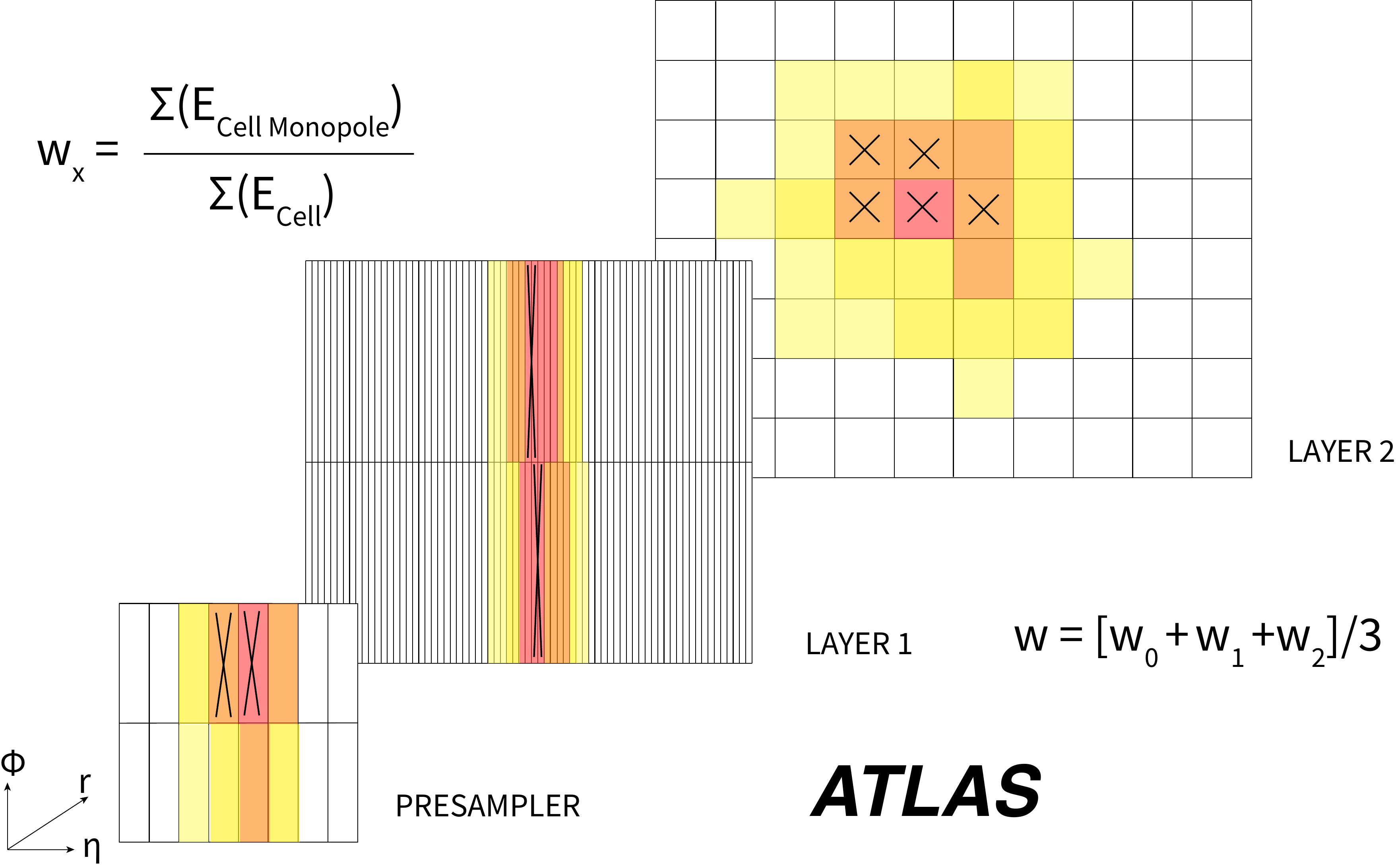}
\caption{Illustration of example energy depositions in each layer of the EM calorimeter. Cells with higher energy deposits are shown in red, while lower energy deposits are yellow. Crosses represent the cells selected for their high-energy deposition for each layer for the computation of the $w$ variable.}\label{fig:wvar}
\end{figure}

\begin{figure}[h!]
\centering
\includegraphics[width=0.8\linewidth,keepaspectratio=true]{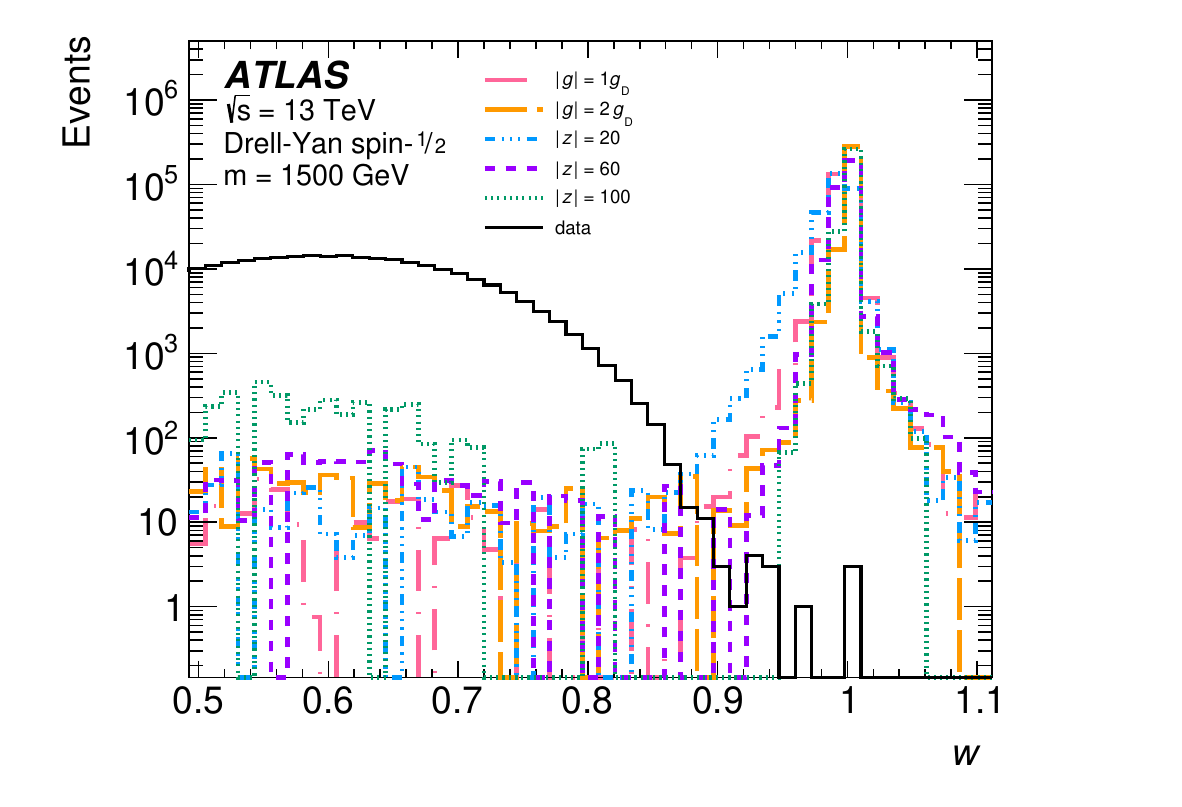}
\caption{Distribution of the $w$ variable for events that passed the HIPTRT trigger and the criteria $|\eta| < 1.375$ and $N_\mathrm{HT}>30$. Only the TRT road candidate with the highest $f_\text{HT}$ value in each event is included in the plot. The black distribution corresponds to 20\% of the Run~2 data and the coloured distributions correspond to the mass 1500~GeV Drell-Yan spin-\textonehalf\ signal samples for various charges. All MC distributions have been normalized to the number of events in the data.}\label{fig:w1d}
\end{figure}

Two powerful variables distinguish HIP signal events from background. The fraction of all the TRT hits in the road (wedge) that exceed the high threshold, $f_\mathrm{HT}$, is the first signal-discriminating variable, shown in Figure~\ref{fig:fht1d}. In the case where there are multiple HIP candidates in an event (this occurs for about 10\% of the selected data events), the candidate with the highest $f_\mathrm{HT}$ value is selected.
The second variable, $w$, reflects the lateral dispersion of the EM cluster, which is narrow for HIPs, since they do not induce an EM shower.
Three variables, $w_0$, $w_1$ and $w_2$, are defined as the fractions of EM cluster energy ($E_i$) contained in the two most energetic cells in the presampler, the four most energetic cells in the EM1 layer, and the five most energetic cells in the EM2 layer, respectively, as depicted in Figure~\ref{fig:wvar}.
The variable $w$, shown in Figure~\ref{fig:w1d}, is defined as the average of all $w_i$ for which the energy $E_i$ exceeds \SI{10}{\giga\ev} in each of the presampler and EM1 layers, and \SI{5}{\giga\ev} in the EM2 layer. In addition, it is required that
one of $E_0$ and $E_1$ exceeds 10~GeV,
as a HIP must pass through the presampler and EM1 layer to reach the EM2 layer and deposit energy there.
 
HIP candidates are identified by the signal discriminating criteria $f_\mathrm{HT} \ge 0.77$ and $w \ge 0.93$.
Optimal values of $f_\mathrm{HT}$ and $w$ were obtained by comparing the signal efficiency for each DY spin-\textonehalf\ MC sample to the rejection of background, represented by a sample of one million pseudodata events prepared by sampling the 1D $f_\mathrm{HT}$ and $w$ distributions in data and randomly pairing the values. Then, the weighted average of all $f_\mathrm{HT}$ and $w$ values found for the different DY spin-\textonehalf\ MC samples was computed, where the weight was given by the efficiency for passing all the selection criteria before the discriminating variable cuts.
 
The method used to extrapolate the selection efficiencies of DY spin-0 HIPs and PF HIPs of both spins is as follows. Given the $E^\text{kin}_\text{T}$ and $\eta$ values for each generated HIP, the corresponding bin in the appropriate efficiency map is identified and the efficiency is extracted. A random number $r$ between 0 and 1 is generated and compared with this probability; if $r$ is less than or equal to the efficiency, then the particle is considered to have passed the selection criteria. This process of extrapolating the selection efficiencies from the generator-level kinematics and the efficiency maps was validated using the fully simulated DY spin-\textonehalf\ samples.
 
\begin{figure}[ht]
\centering
\begin{minipage}{0.49\linewidth}\subfloat{\includegraphics[width=0.98\linewidth,keepaspectratio=true]{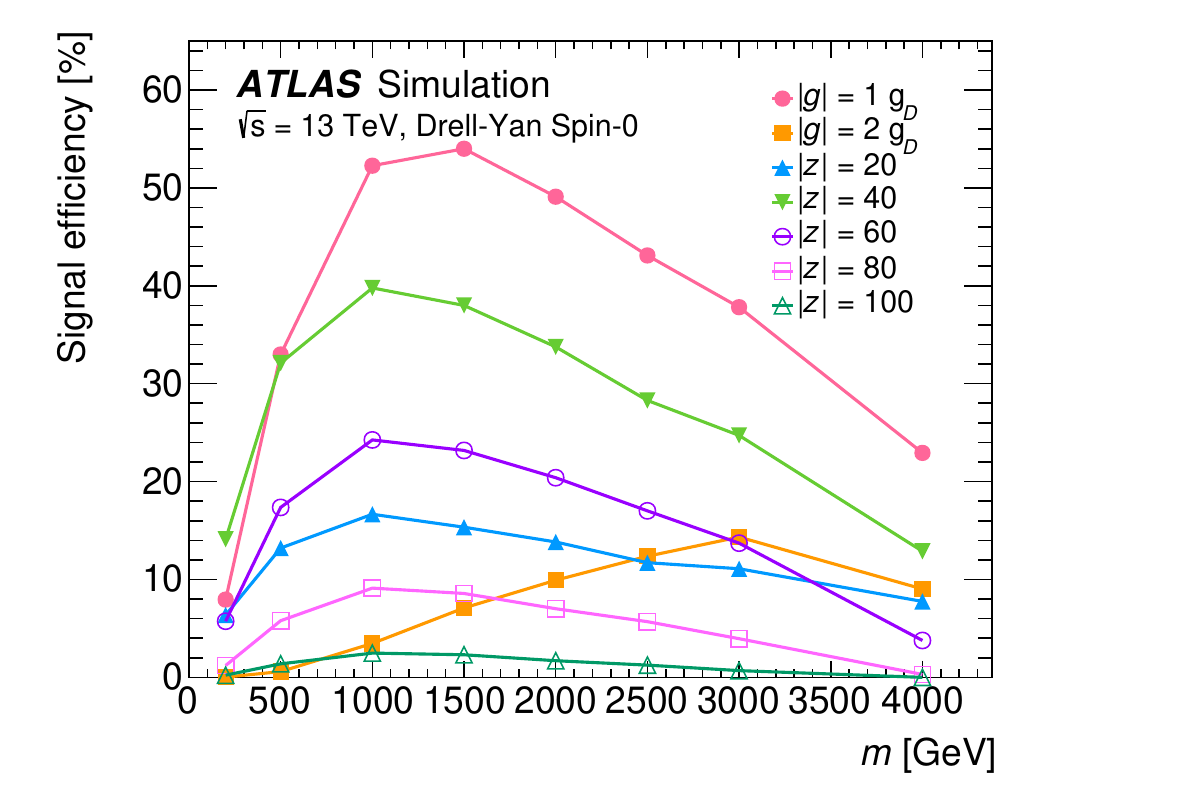}}\end{minipage}
\begin{minipage}{0.49\linewidth}\subfloat{\includegraphics[width=0.98\linewidth,keepaspectratio=true]{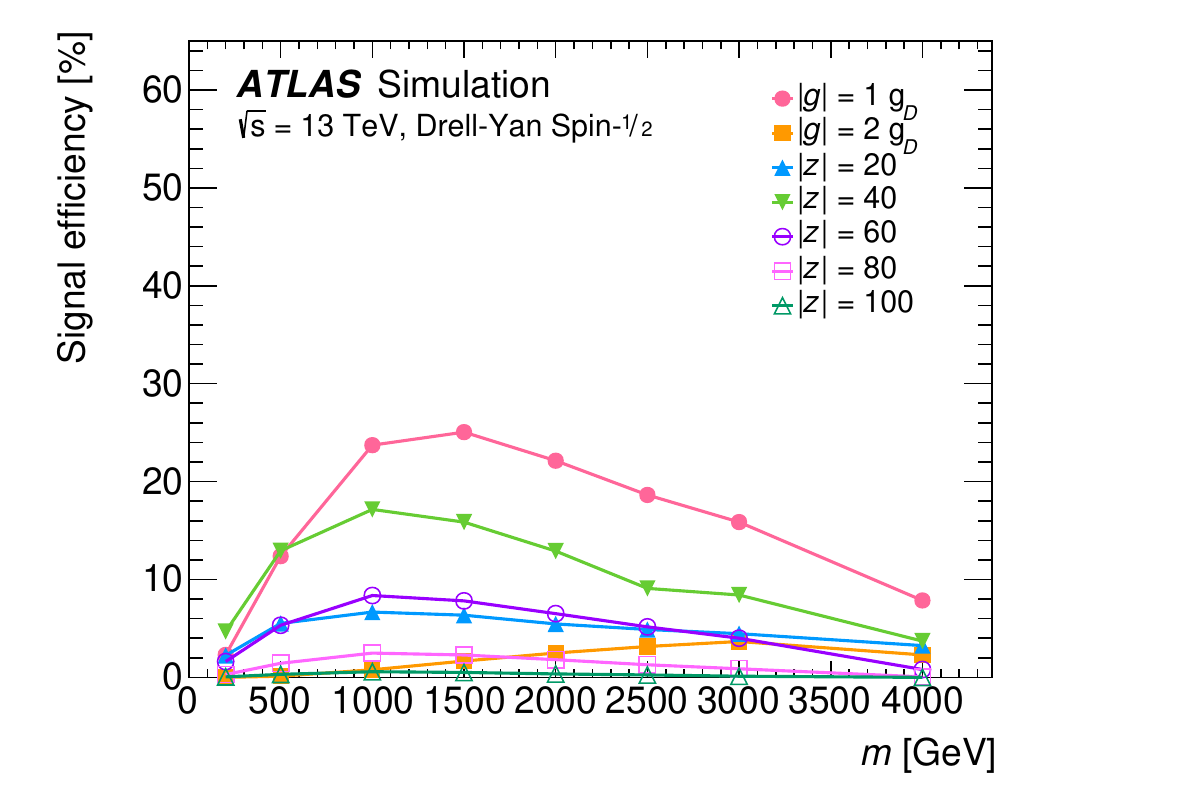}}\end{minipage}
\begin{minipage}{0.49\linewidth}\subfloat{\includegraphics[width=0.98\linewidth,keepaspectratio=true]{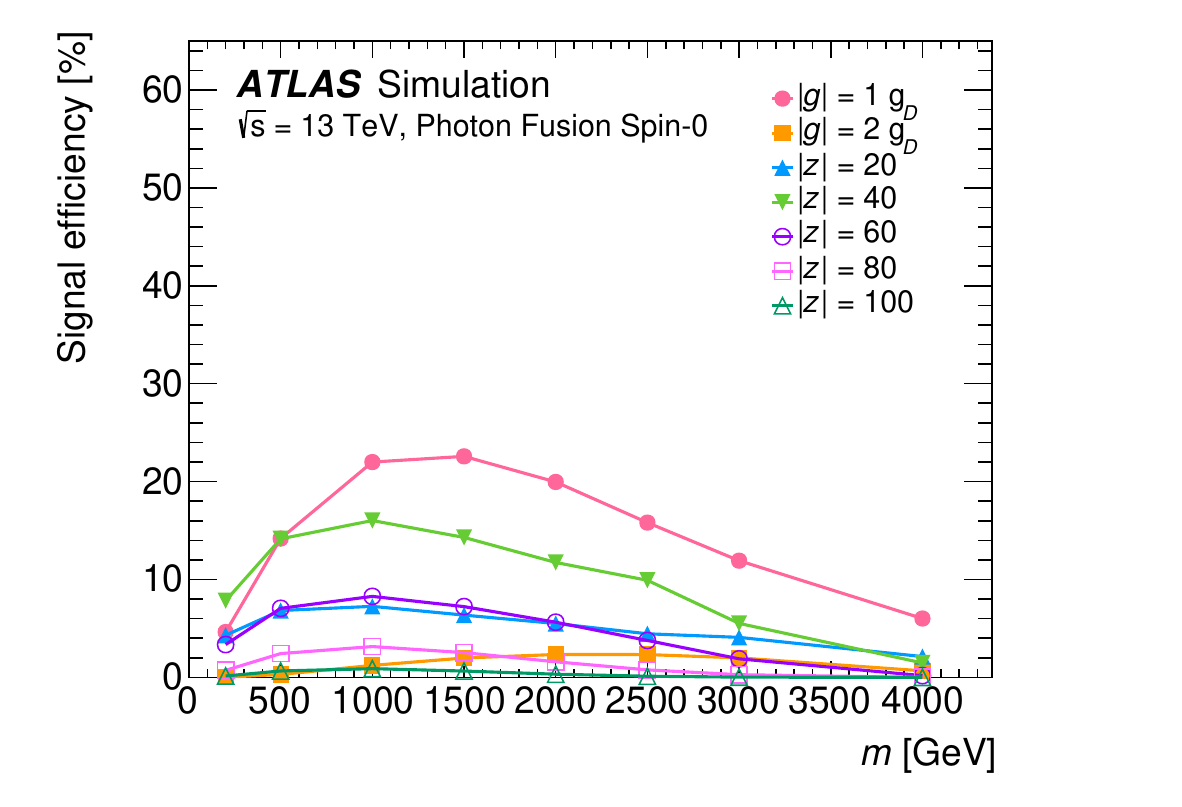}}\end{minipage}
\begin{minipage}{0.49\linewidth}\subfloat{\includegraphics[width=0.98\linewidth,keepaspectratio=true]{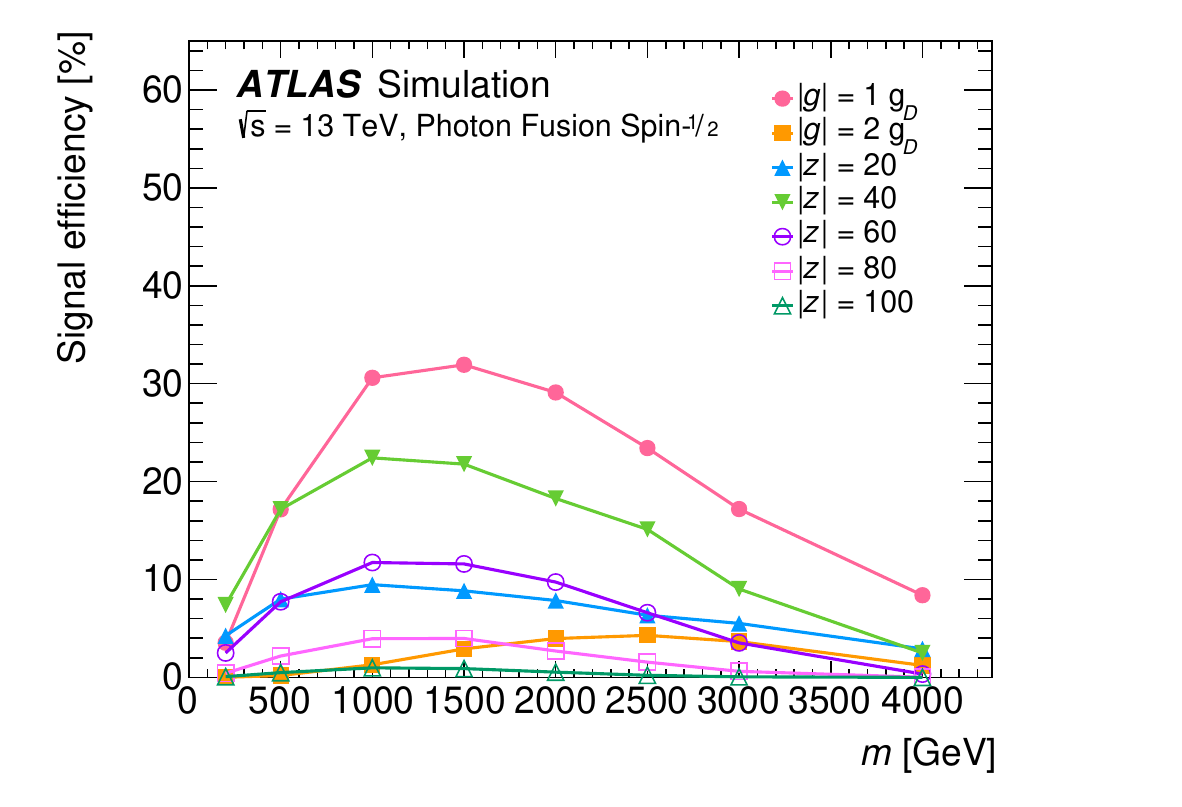}}\end{minipage}
\caption{Selection efficiencies for Drell-Yan (top) and photon-fusion
(bottom) pair-produced monopoles and HECOs, for spin-0 (left) and
spin-\textonehalf\ (right).
Only Drell-Yan spin-\textonehalf\ efficiencies come from fully simulated samples. Efficiencies for all other considered mechanisms are extrapolated from efficiency maps.}\label{fig:efficiencies}
\end{figure}

The signal efficiencies for a given mass, charge and spin for each production mode are shown in Figure~\ref{fig:efficiencies}. The selection efficiencies for spin-0 DY HIPs are higher than those for spin-\textonehalf\ DY HIPs because the former have more central $|\eta|$ and harder kinetic energy distributions. On the other hand, the selection efficiencies for spin-0 PF HIPs are lower than those for spin-\textonehalf\ PF HIPs, which have harder kinetic energy distributions. The HIP efficiencies are strongly correlated with charge and mass, with the selection of HIPs in the intermediate charge and mass ranges being the most efficient. In general, the high ionization of the higher charge HIPs is such that they stop before the EM calorimeter, whereas the lower charge HIPs may reach the hadronic calorimeter such that they are vetoed by the first-level trigger. The lower mass HIPs have lower efficiency because they have lower kinetic energies and may fail the first-level trigger, whereas higher mass HIPs may fail the trigger due to late arrival at the EM calorimeter.
 
\section{Background Estimate}
\label{sec:bkg}
\begin{figure}
\centering
\includegraphics[width=0.8\linewidth,keepaspectratio=true]{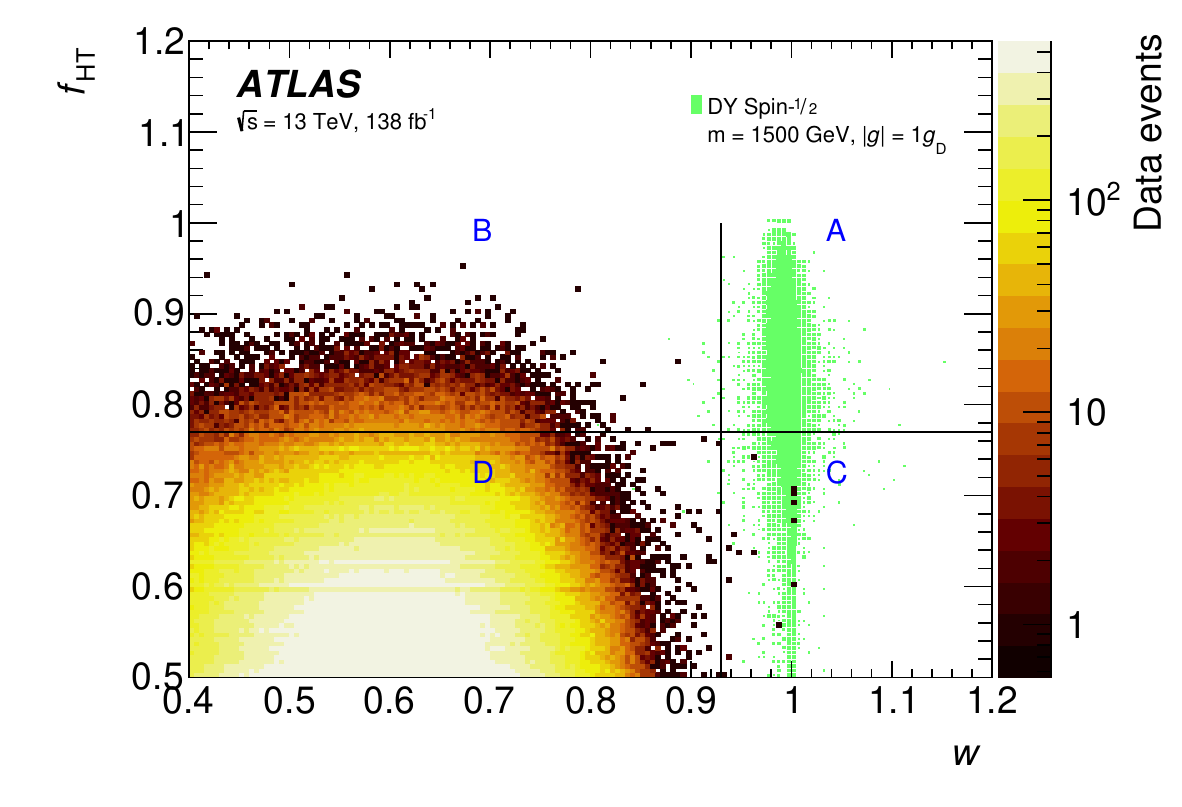}
\caption{Two-dimensional distribution of the discriminators $f_\mathrm{HT}$ and $w$ for the data and a representative signal sample (green). The signal (A) and control (B, C and D) regions are shown.}\label{fig:abcd}
\end{figure}

Random combinations of rare processes, such as overlapping charged particles in multijet events and superpositions of high-energy electrons, can occasionally produce high $f_\text{HT}$ and high $w$ values.
Considering how rare such events are, it
is not practical to simulate a statistically significant MC sample. Instead, a data-driven ABCD background estimation approach is used.
The 2D distribution of $f_\text{HT}$ versus $w$ for the HIP candidates, shown in Figure~\ref{fig:abcd}, is divided into four regions: signal region A ($f_\mathrm{HT} \geq 0.77$ and $w \geq 0.93$), which contains most of the signal events, and three neighboring control regions B ($f_\mathrm{HT} \geq 0.77$ and $0.4 \leq w < 0.93$), C ($0.5 \leq f_\mathrm{HT} < 0.77$ and $w \geq 0.93$) and D ($0.5 \leq f_\mathrm{HT} < 0.77$ and $0.4 \leq w < 0.93$).
If there is no correlation between the two discriminating variables, the estimate of expected background events in region A is calculated as $N_A^\text{exp}=N_BN_C/N_D$,
where $N_B$, $N_C$ and $N_D$ are the numbers of events in regions B, C and D, respectively, as shown in Table~\ref{tbl:yields}.
\begin{table}[tb]
\caption{Data event yields in the signal region A and the three control regions, B, C and D}  \label{tbl:yields}
\begin{center}
\begin{tabular}{ccccc}
\hline
\hline
region & A & B & C & D \\
\hline
events & 0 & 9847 & 15 & 986500 \\
\hline
\hline
\end{tabular}
\end{center}
\end{table}
 
The ratio, $N_B/N_D$, called the transfer factor, is observed to fluctuate by up to 50\% in bins of $w$. This is due to a mutual correlation of both variables, $f_\mathrm{HT}$ and $w$, with $\eta$, reflecting features in the detector structure, such as the barrel-end-cap transition region.
The small effect of the non-uniformity on the search results can be evaluated by
splitting the data into three $|\eta|$ ranges, resulting in a total of 12 orthogonal regions or bins on the ABCD plane.
The mean transfer factor for each $|\eta|$ range (see Table~\ref{tbl:TF_eta}) is determined by fitting a constant to the $N_B/N_D$ versus $w$ distribution and
used to obtain a background estimate $N_A^\text{exp}$ for each $|\eta|$-sliced ABCD plane.
This gives rise to a relative systematic uncertainty of 30\%, which is assigned to the final background estimate.
\begin{table}[h]
\begin{center}
\caption{Mean transfer factor for different $|\eta|$ regions. The variations in the statistical uncertainties reflect the $\eta$ dependence of the selection efficiency. The root mean square of the constant fit to the $N_B/N_D$ versus $w$ distribution is assigned as the systematic uncertainty. Transfer factor fluctuations due to the low selection efficiency in the $|\eta|<0.32$ region lead to a larger systematic uncertainty than in the other regions.}
\label{tbl:TF_eta}
\begin{tabular}{cc}
\hline
\hline
$|\eta|$ range & mean transfer factor \\
\hline
0 - 0.32     & $0.0087 \pm 0.0014\textrm{(stat.)} \pm 0.0066\textrm{(syst.)} $ \\
0.32 - 0.82  & $0.00125 \pm 0.00005\textrm{(stat.)} \pm 0.00021\textrm{(syst.)}$ \\
0.82 - 1.375 & $0.0158 \pm 0.0001\textrm{(stat.)} \pm 0.0027\textrm{(syst.)}$ \\
\hline
\hline
\end{tabular}
\end{center}
\end{table}
 
To ensure that the ABCD method yields a prediction of background events close to the actual number in
the signal region, the method is validated as follows. Control region B is divided into validation regions B$^\prime$ ($w<0.665$) and B$^{\prime\prime}$ ($w\ge0.665$), containing 7684 and 2163 events, respectively. Validation regions D$^\prime$ and D$^{\prime\prime}$, containing 754~040 and 232~460 events, respectively, are similarly obtained from control region D.
The background estimate $N_{\text{B}^{\prime\prime}}^\text{exp}=2368$ for the number of events in region B$^{\prime\prime}$, found by computing $N_{\text{B}^{\prime\prime}}^\text{exp}=N_{\text{B}^\prime} N_{\text{D}^{\prime\prime}}/N_{\text{D}^\prime}$,
is in reasonable agreement with the observed number of data events in that region.
 
The HIP signal leakage into the control regions B, C, and D, as seen in Figure~\ref{fig:abcd}, could bias the background estimate and, hence, is considered for the fully simulated spin-\textonehalf\ DY pair-production MC samples.
A background estimate accounting for signal leakage can be computed based on a simultaneous likelihood fit to signal and background
of the ABCD plane. The inputs to this fit are the background estimated by the aforementioned method for the full $\eta$ range,
the number of data events in regions B, C and D, with a hypothesis of the number of events in region A,
and the fraction of signal events in all regions. 
Thus, a likelihood function is constructed as a product of Poisson probabilities for all four regions, constrained by
Gaussian distributions that take into account as nuisance parameters the signal efficiency
systematic uncertainties, described in Section~\ref{sec:uncert}, and the MC statistical uncertainties.
By maximizing the likelihood function in the fit to the observed data, the background is estimated to be $N_A^\text{exp}=0.15\pm0.04 (\text{stat.}) \pm0.05 (\text{syst.})$, where the systematic uncertainty is that derived from the $|\eta|$-sliced ABCD method and accounts for the transfer factor non-uniformity. This estimate is in agreement with the background estimate of the ABCD method.
 
The signal leakage of the spin-0 DY pair-produced HIPs and the PF pair-produced HIPs of either spin is not considered because it cannot be estimated using the extrapolation method described in Section~\ref{sec:signal}. The reason is that most signal events appear in the signal region A; consequently, single-particle samples many times larger than the existing samples would be needed to sufficiently populate control region "efficiency" maps.
Neglecting signal leakage does not introduce any significant bias into the limits for the alternative models presented in Section~\ref{sec:results}, since an average relative discrepancy of less than 1\% was found when comparing the limits computed considering signal leakage into the control regions to the limits obtained neglecting signal leakage for the DY spin-\textonehalf\ production mode.

\section{Systematic Uncertainties}
\label{sec:uncert}
 
There are various sources of uncertainty that arise in the computation of the signal selection efficiency,
most of which are induced by possible imperfections of the HIP signal simulation.
Uncertainties in the signal efficiencies are evaluated for each HIP charge, mass, spin and production mode, but only average relative uncertainties are reported here.
Uncertainties evaluated in one direction are assumed to be symmetric.

There are several selection efficiency uncertainties associated with the simulation of the ionization process. These are quantified by simulating an alternative signal sample with modified parameters.
The detector material mismodeling leads to the largest uncertainty, which inherits the dependence on the square of the charge from the ionization stopping power.
Increasing the material in the ID and the EM calorimeter by one standard deviation (a 7.5\% increase of the ID material~\cite{PERF-2015-07}, a 10\% increase of the ID services material~\cite{PERF-2015-07}, and a 5\% increase in the calorimeter presampler material) yields a relative uncertainty in the selection efficiency of 4\% for $1g_\textrm{D}$ and 18\% for $2g_\textrm{D}$ monopoles of mass 1500~GeV. On average across all monopole and HECO charges, the relative uncertainty is 9\%.
The $\delta$-ray production is dictated by the $dE/dx$ formulas for ionization by monopoles and HECOs, which have theoretical uncertainties of about 3\%~\cite{Ahlen1978,Ahlen1980} in the kinematic range considered in this analysis.
By reducing the $\delta$-ray production by 3\% in the simulation, an average relative uncertainty of 2\% is estimated.
Another uncertainty is related to electron-ion recombination in the LAr EM calorimeter, which reduces the energy recorded compared to the energy deposited by a particle. While Birks’ law models this effect as a function of $dE/dx$~\cite{Birks1952}, it overestimates the recombination effects at high $dE/dx$. A correction to Birks’ law for HIPs, as described in Ref.~\cite{HIPBirks}, is implemented in the simulation. By varying this correction within its uncertainties, an average relative uncertainty of 8\% is assigned.

The TRT occupancy (the fraction of TRT straws with hits in any event) is slightly underestimated in the simulation.
This mismodeling affects the fraction of TRT HT hits used as a discriminating variable in the trigger and offline selections.
Scaling the number of LT hits (those likely to come from pileup as opposed to HIPs or associated $\delta$-rays) in each event by pileup-dependent ratios of TRT occupancy between data and simulation results, on average, in a relative systematic uncertainty of 2\% in the selection efficiency. Similarly, a relative systematic uncertainty of 2\% accounting for potential mismodeling of pileup is estimated with variations of the nominal pileup distribution within its uncertainty.
 
Slow-moving HIPs ($\beta < 0.37$) arrive at the EM calorimeter with a time delay of more than 10~ns with respect to the HIP production time. As a result, the first-level trigger may associate the calorimeter cluster with the next bunch crossing, leading to a reduction of the HIP selection efficiency by the HLT. However, this effect is not accurately simulated,
such that the trigger efficiency for slow-moving HIPs is overestimated.
Therefore, the trigger efficiency loss is determined by rejecting the slow-moving HIPs at truth level, leading to a relative systematic uncertainty of 2\% on average in the selection efficiency.

Finally, by comparing the efficiencies extrapolated from single-particle efficiency maps to those obtained using fully simulated MC spin-\textonehalf\ samples, an average relative systematic uncertainty of 1\% in the selection efficiency is found for imprecise extrapolation of Drell-Yan spin-0 and photon-fusion efficiencies.

For each HIP charge, mass, spin and production mode, the systematic uncertainties described above are added in quadrature to obtain a final uncertainty on the signal efficiency.
The limit-setting framework presented in Section~\ref{sec:results}
also takes into account the 30\% relative systematic uncertainty on the background estimate described in Section~\ref{sec:bkg} and the integrated luminosity uncertainty, which is 0.83\%, estimated following the methods discussed in Ref.~\cite{Run2lumi}.

\section{Upper Cross-section and Lower Mass Limits}
\label{sec:results}
\begin{figure}[ht]
\centering
\begin{minipage}{0.49\linewidth}\subfloat{\includegraphics[width=0.98\linewidth,keepaspectratio=true]{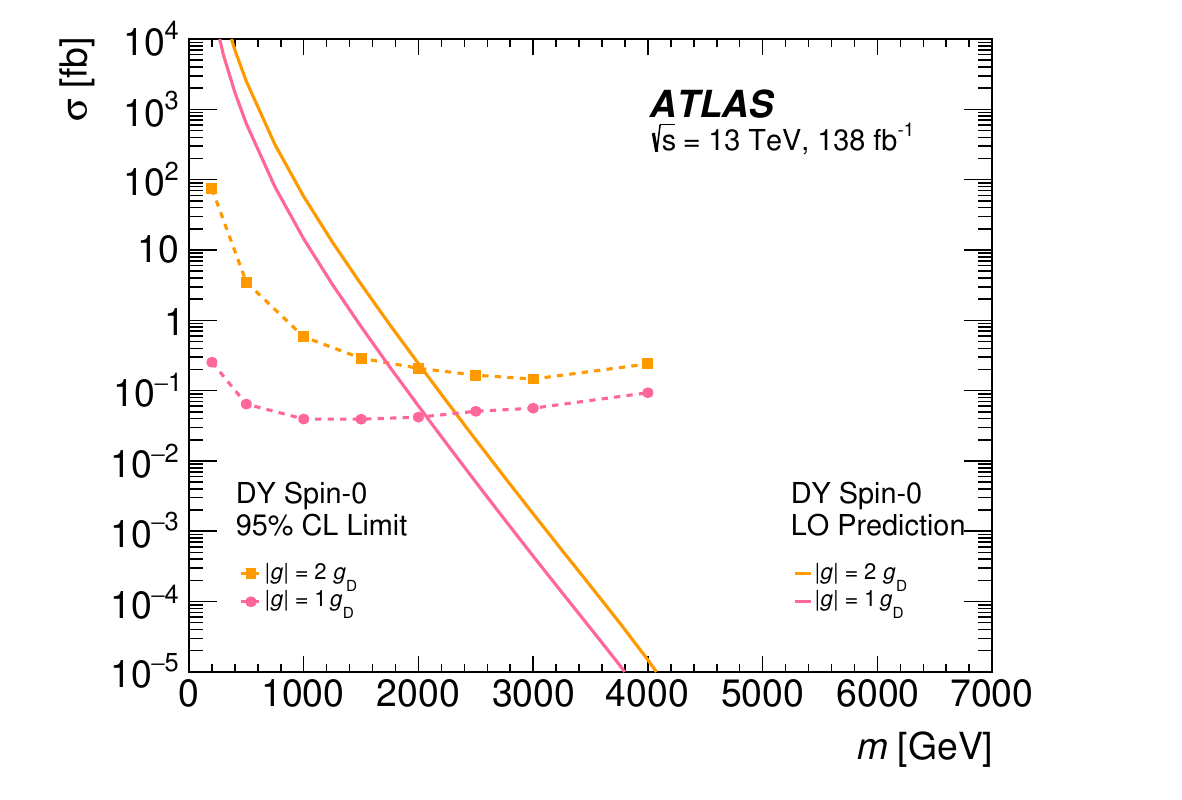}}\end{minipage}
\begin{minipage}{0.49\linewidth}\subfloat{\includegraphics[width=0.98\linewidth,keepaspectratio=true]{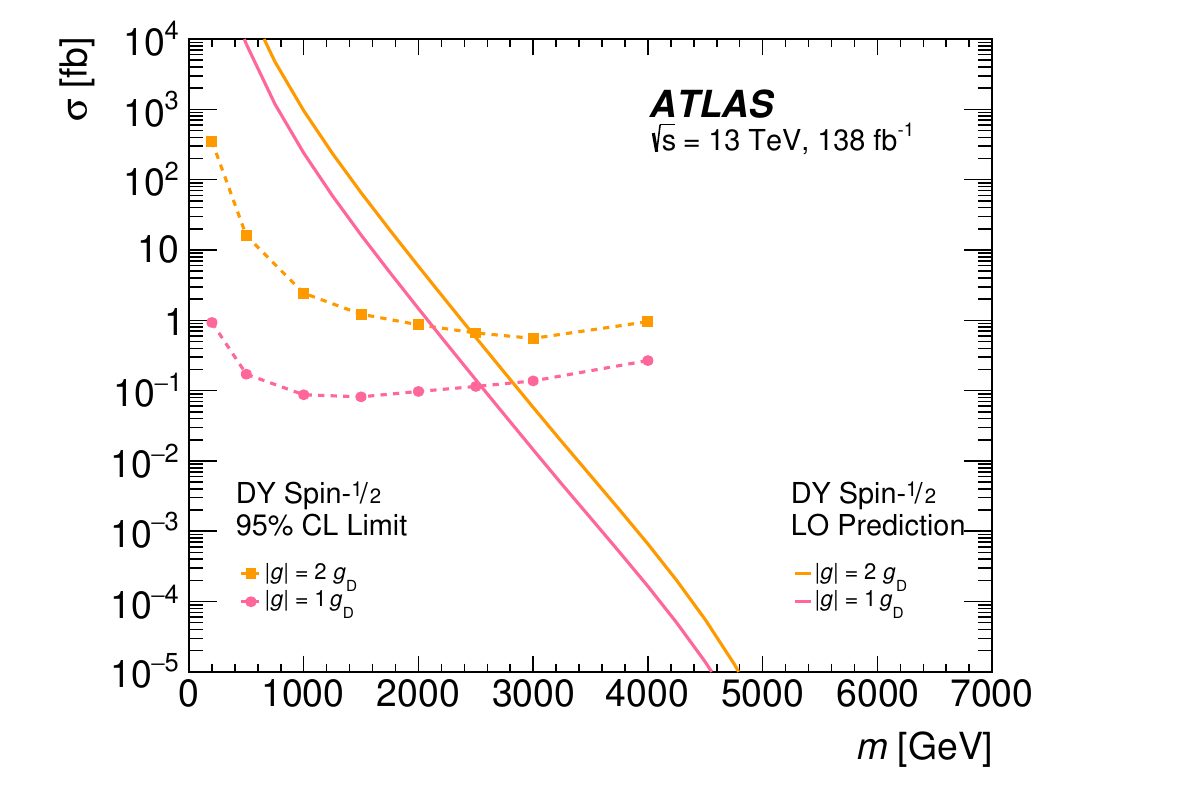}}\end{minipage}
\begin{minipage}{0.49\linewidth}\subfloat{\includegraphics[width=0.98\linewidth,keepaspectratio=true]{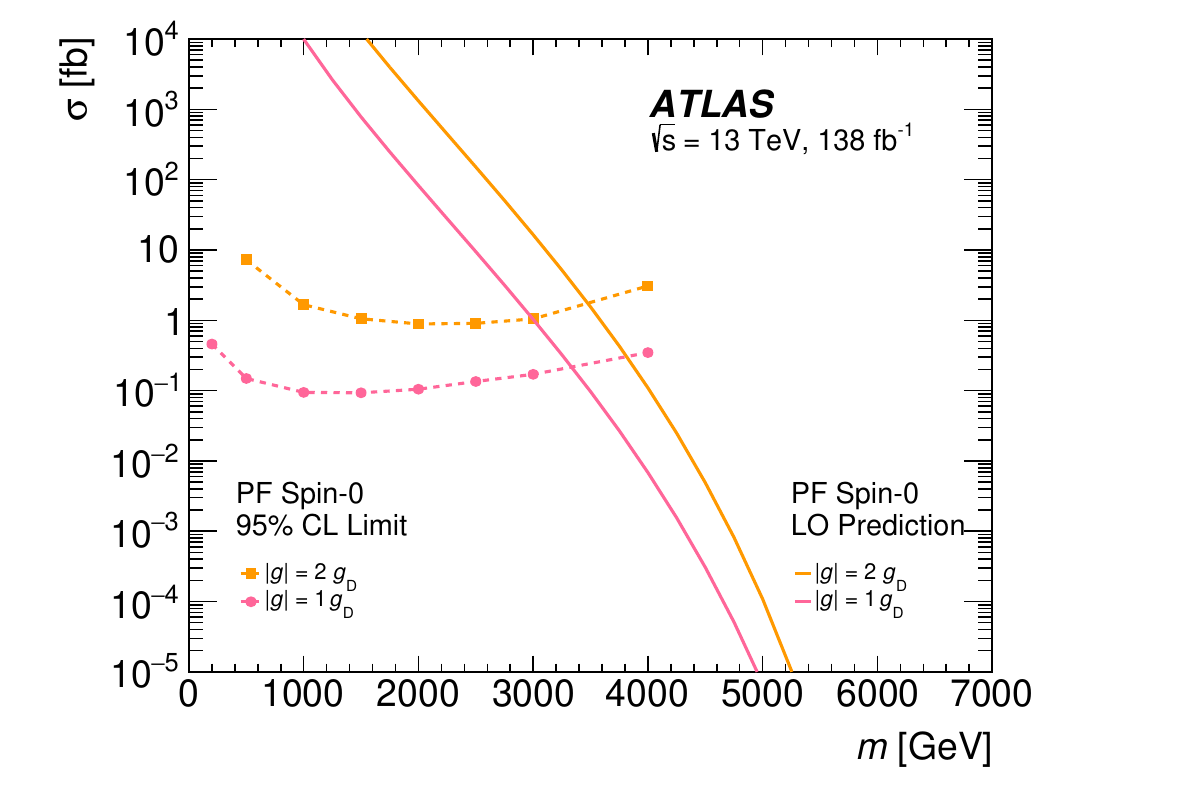}}\end{minipage}
\begin{minipage}{0.49\linewidth}\subfloat{\includegraphics[width=0.98\linewidth,keepaspectratio=true]{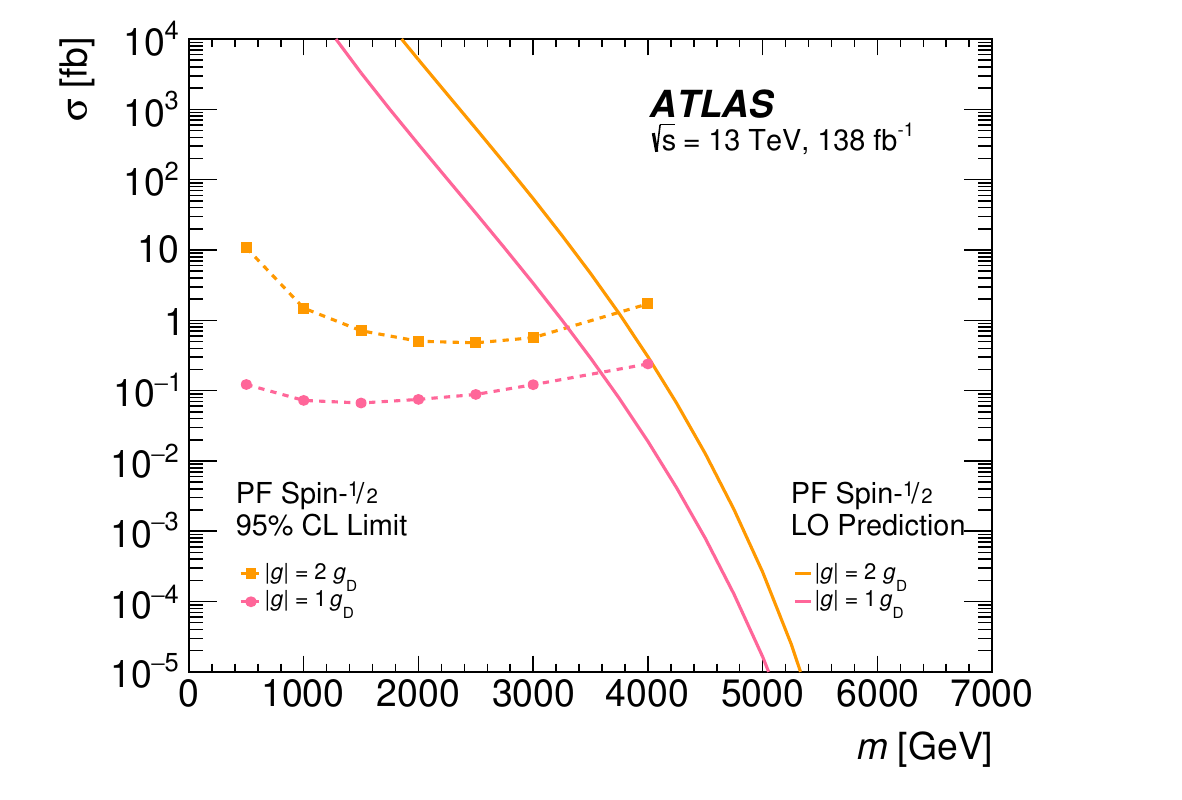}}\end{minipage}
\caption{Observed 95\% CL upper limits on the cross section for all masses and charges of Drell-Yan (top) and photon-fusion (bottom) pair-produced monopoles for spin-0 (left) and spin-\textonehalf\ (right). Mass points for which the selection efficiency is below 0.1\% are not shown. The solid lines are the theoretical  leading-order cross sections as a function of mass.}\label{fig:lim_MONO}
\end{figure}

\begin{figure}[ht]
\centering
\begin{minipage}{0.49\linewidth}\subfloat{\includegraphics[width=0.98\linewidth,keepaspectratio=true]{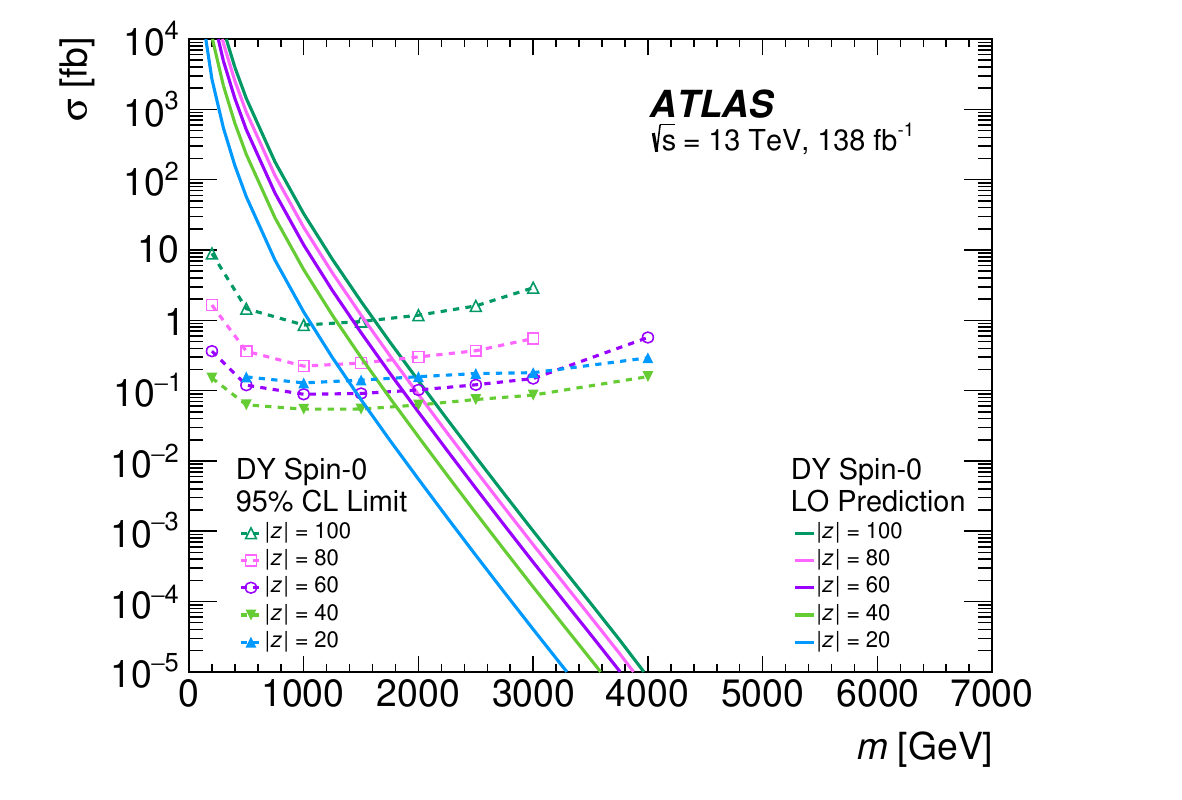}}\end{minipage}
\begin{minipage}{0.49\linewidth}\subfloat{\includegraphics[width=0.98\linewidth,keepaspectratio=true]{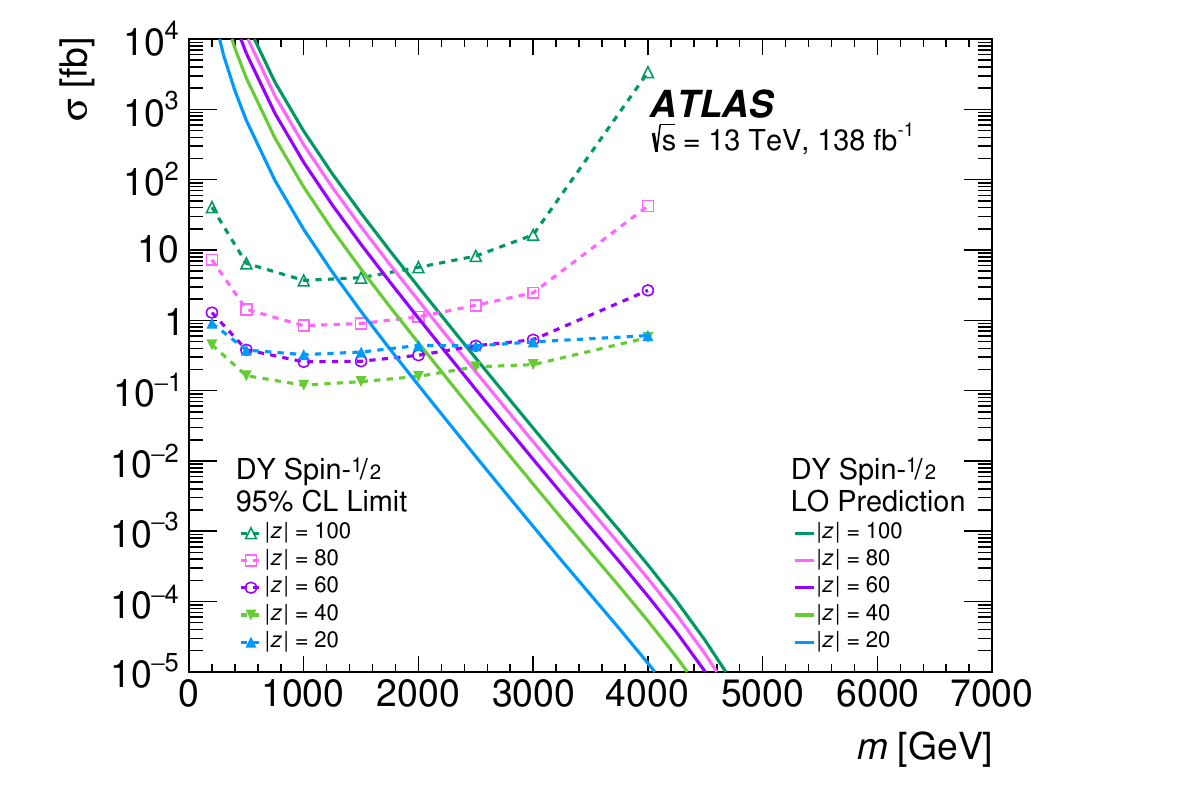}}\end{minipage}
\begin{minipage}{0.49\linewidth}\subfloat{\includegraphics[width=0.98\linewidth,keepaspectratio=true]{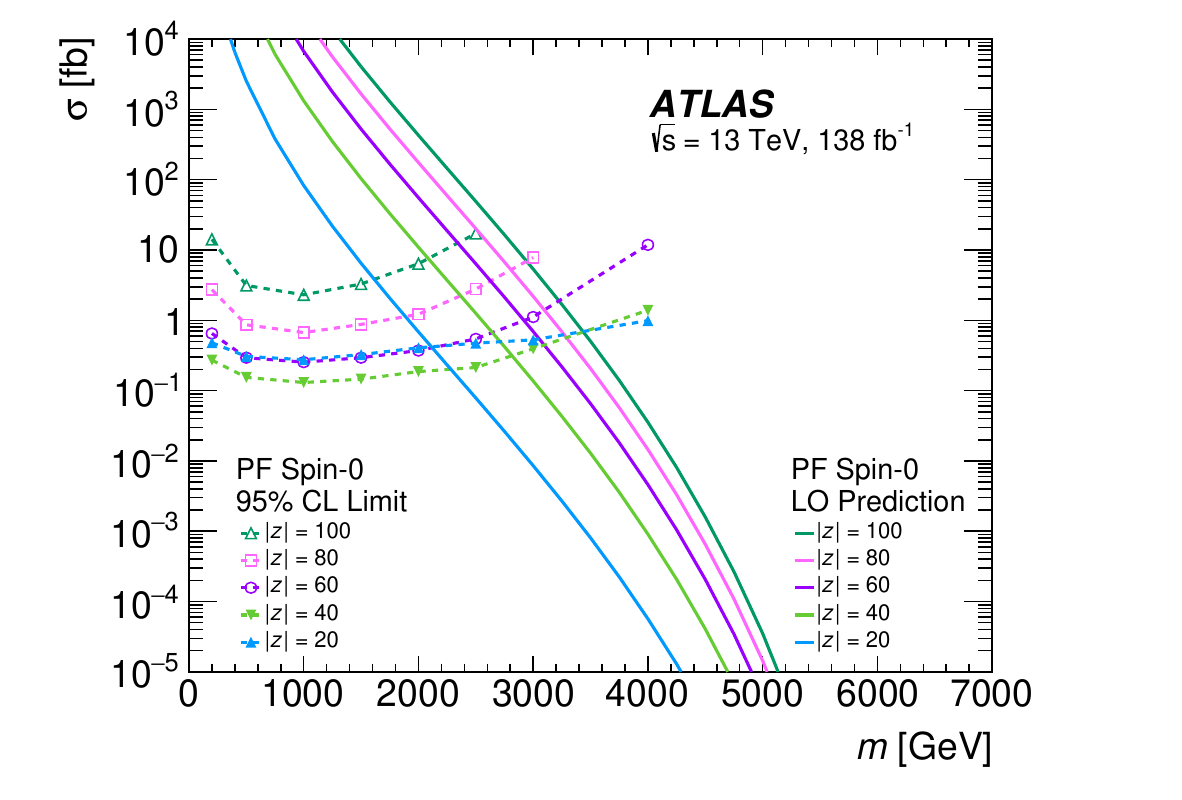}}\end{minipage}
\begin{minipage}{0.49\linewidth}\subfloat{\includegraphics[width=0.98\linewidth,keepaspectratio=true]{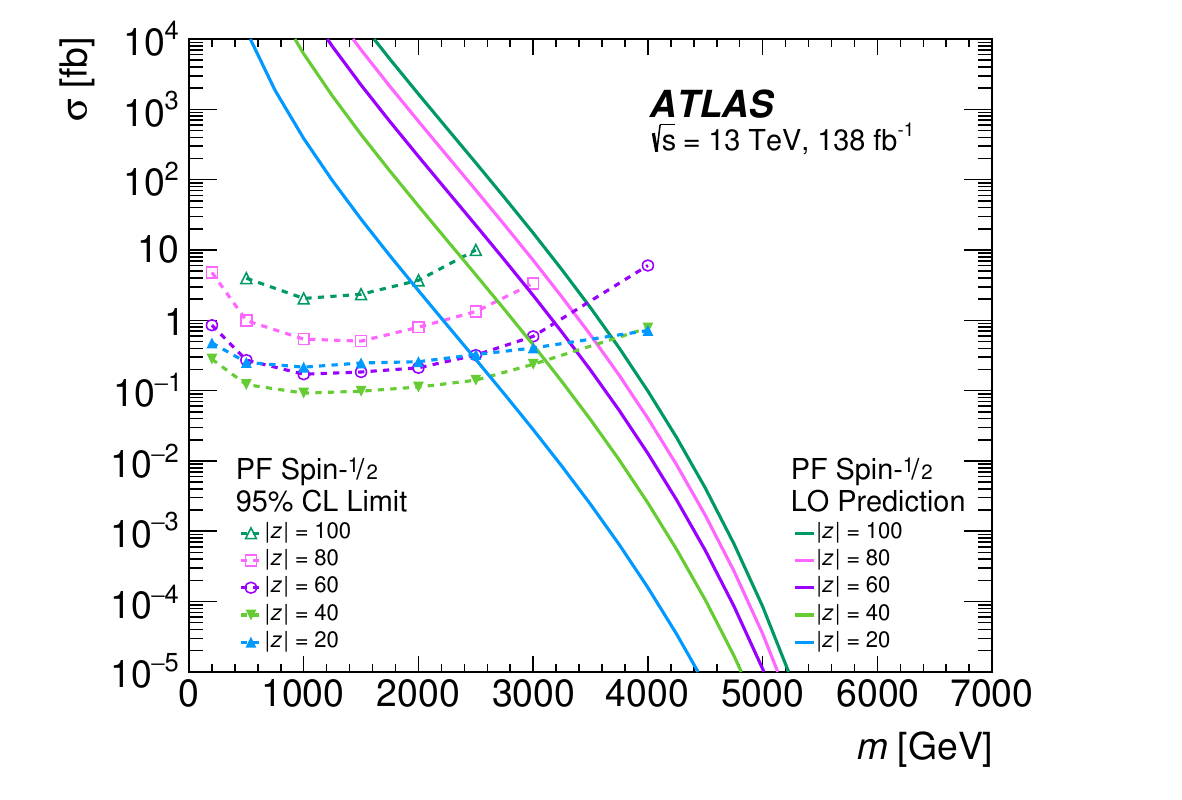}}\end{minipage}
\caption{Observed 95\% CL upper limits on the cross section for all masses and charges of Drell-Yan (top) and photon-fusion (bottom) pair-produced HECOs for spin-0 (left) and spin-\textonehalf\ (right). Mass points for which the selection efficiency is below 0.1\% are not shown. The solid lines are the theoretical leading-order cross sections as a function of mass.}\label{fig:lim_HECO}
\end{figure}

\begin{figure}[h!]
\centering
\includegraphics[width=0.8\linewidth,keepaspectratio=true]{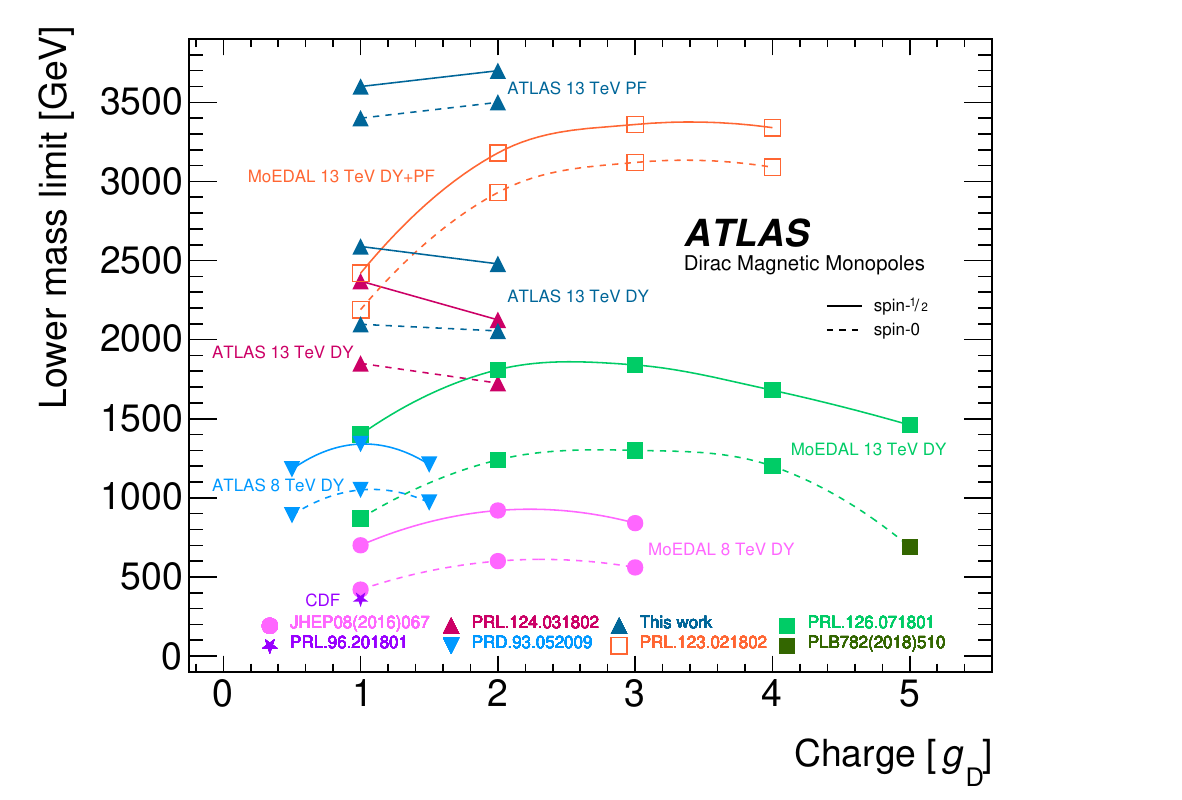}
\caption{Comparison of the lower mass limits obtained by LHC searches in Run~1 and Run~2 $pp$ collisions for Drell-Yan and photon-fusion pair-produced magnetic monopoles. A Tevatron measurement by CDF in $p\bar{p}$ collisions is also shown. The dashed and solid lines represent spin-0 and spin-\textonehalf\ measurements, respectively.}\label{fig:masscomp}
\end{figure}

Consistent with the background expectation, no event is observed in the HIP signal region. Therefore, exclusion limits are set on all four signal models at 95\% confidence level (CL) using the CL$_S$~\cite{Read:2002hq} frequentist framework implemented in RooStats~\cite{RooFitConf}.
The cross-section limits are obtained exploiting selection efficiencies and their corresponding uncertainties for each HIP signal sample, the systematic uncertainty on the background estimate, and the integrated luminosity uncertainty. A toys-based approach, with 5000 pseudoexperiments, is used to approximate the test-statistic distribution for each considered signal model.
The likelihood function for DY spin-\textonehalf\ HIPs considers the signal leakage in the control regions, whereas only signal-region information is used for DY spin-0 pair-produced HIPs and PF pair-produced HIPs of both spins.
Figures~\ref{fig:lim_MONO} and \ref{fig:lim_HECO} show the observed 95\% CL upper limits on the production cross sections, as a function of the HIP mass. Cross section limits for which the selection efficiency is below 0.1\% are not reported, since the limited statistics preclude the evaluation of reliable systematic uncertainties in those cases.
Lower mass limits derived for each production mode are shown in Table~\ref{tbl:masslim}. In cases where the cross section upper limit curve does not intersect the production cross section curve, the mass limit is set to the highest mass for which a cross section upper limit was determined.
\begin{table}[h]
\caption{Lower limits on the mass of magnetic monopoles and HECOs (in TeV) at
95\% confidence level in models of spin-0 and spin-\textonehalf\ Drell-Yan (DY) and photon-fusion (PF) pair production.}\label{tbl:masslim}
\begin{center}
\begin{tabular}{lccccccc}
\hline
\hline
& \multicolumn{7}{c}{95\% CL lower limits on the mass of HIPs [TeV]}\\
& $|g|=1g_\text{D}$ & $|g|=2g_\text{D}$ &
$|z|=20$ & $|z|=40$ & $|z|=60$ & $|z|=80$ & $|z|=100$ \\
\hline
DY spin-0 & 2.1 & 2.1 & 1.4 & 1.8 & 1.9 & 1.8 & 1.7 \\
DY spin-\textonehalf\ & 2.6 & 2.5 & 1.8 & 2.2 & 2.2 & 2.1 & 1.9 \\
PF spin-0 & 3.4 & 3.5 & 2.1 & 2.8 & 2.9 & 2.8 & 2.5 \\
PF spin-\textonehalf\ & 3.6 & 3.7 & 2.5 & 3.1 & 3.1 & 3.0  & 2.5 \\
\hline
\hline
\end{tabular}
\end{center}
\end{table}
The photon-fusion mechanism provides more stringent mass limits than the Drell-Yan mechanism, as it has a higher predicted cross section. Given that the predicted cross sections can only be calculated at leading order, these mass limits primarily serve as benchmarks for comparison with other experiments, as shown in Figure~\ref{fig:masscomp}. The ATLAS cross-section limits for Drell-Yan production of HECOs in the range $20\le|z|\le100$ are several orders of magnitude better than those reported by MoEDAL in Ref.~\cite{MoEDAL2021HIP}, which considers \SI{8}{\tera\ev} $pp$ collision data. While MoEDAL is the only experiment sensitive to magnetic charges in the range $3g_\textrm{D} \le |g| \le 5g_\textrm{D}$~\cite{MoEDAL2016a,MoEDAL2017a,MoEDAL2018,MoEDAL2019,MoEDAL2021Schwinger}, the present ATLAS search is able to set significantly better cross-section limits (one to two orders of magnitude) for $1g_\textrm{D}$ and $2g_\textrm{D}$.
 
\FloatBarrier
 
\section{Conclusions}
\label{sec:conclusions}
 
This paper presents a search for spin-0 and spin-\textonehalf\ magnetic monopoles and HECOs produced via the Drell-Yan and photon-fusion mechanisms using a dataset of \SI{138}{\per\fb} of 13~TeV proton-proton collisions collected by the ATLAS detector between 2015 and 2018 during Run~2.
Upper cross-section limits and lower mass limits computed at 95\% confidence level are presented.
The search improves by approximately a factor of three the cross-section limits on the Drell-Yan production of magnetic monopoles with magnetic charges $1g_\textrm{D}$ and $2g_\textrm{D}$ and HECOs in the range $20\le|z|\le 100$ attained in the 2015/16 dataset alone.
Also, the first ATLAS limits on the photon-fusion pair production mechanism of magnetic monopoles and HECOs are obtained.
 
\section*{Acknowledgements}


We thank CERN for the very successful operation of the LHC, as well as the
support staff from our institutions without whom ATLAS could not be
operated efficiently.
 
We acknowledge the support of
ANPCyT, Argentina;
YerPhI, Armenia;
ARC, Australia;
BMWFW and FWF, Austria;
ANAS, Azerbaijan;
CNPq and FAPESP, Brazil;
NSERC, NRC and CFI, Canada;
CERN;
ANID, Chile;
CAS, MOST and NSFC, China;
Minciencias, Colombia;
MEYS CR, Czech Republic;
DNRF and DNSRC, Denmark;
IN2P3-CNRS and CEA-DRF/IRFU, France;
SRNSFG, Georgia;
BMBF, HGF and MPG, Germany;
GSRI, Greece;
RGC and Hong Kong SAR, China;
ISF and Benoziyo Center, Israel;
INFN, Italy;
MEXT and JSPS, Japan;
CNRST, Morocco;
NWO, Netherlands;
RCN, Norway;
MEiN, Poland;
FCT, Portugal;
MNE/IFA, Romania;
MESTD, Serbia;
MSSR, Slovakia;
ARRS and MIZ\v{S}, Slovenia;
DSI/NRF, South Africa;
MICINN, Spain;
SRC and Wallenberg Foundation, Sweden;
SERI, SNSF and Cantons of Bern and Geneva, Switzerland;
MOST, Taiwan;
TENMAK, T\"urkiye;
STFC, United Kingdom;
DOE and NSF, United States of America.
In addition, individual groups and members have received support from
BCKDF, CANARIE, Compute Canada and CRC, Canada;
PRIMUS 21/SCI/017 and UNCE SCI/013, Czech Republic;
COST, ERC, ERDF, Horizon 2020 and Marie Sk{\l}odowska-Curie Actions, European Union;
Investissements d'Avenir Labex, Investissements d'Avenir Idex and ANR, France;
DFG and AvH Foundation, Germany;
Herakleitos, Thales and Aristeia programmes co-financed by EU-ESF and the Greek NSRF, Greece;
BSF-NSF and MINERVA, Israel;
Norwegian Financial Mechanism 2014-2021, Norway;
NCN and NAWA, Poland;
La Caixa Banking Foundation, CERCA Programme Generalitat de Catalunya and PROMETEO and GenT Programmes Generalitat Valenciana, Spain;
G\"{o}ran Gustafssons Stiftelse, Sweden;
The Royal Society and Leverhulme Trust, United Kingdom.
 
The crucial computing support from all WLCG partners is acknowledged gratefully, in particular from CERN, the ATLAS Tier-1 facilities at TRIUMF (Canada), NDGF (Denmark, Norway, Sweden), CC-IN2P3 (France), KIT/GridKA (Germany), INFN-CNAF (Italy), NL-T1 (Netherlands), PIC (Spain), ASGC (Taiwan), RAL (UK) and BNL (USA), the Tier-2 facilities worldwide and large non-WLCG resource providers. Major contributors of computing resources are listed in Ref.~\cite{ATL-SOFT-PUB-2023-001}.


\clearpage
 
\printbibliography
 
\clearpage
\input{atlas_authlist}

\end{document}

%% file: atlas_authlist.tex
 
\begin{flushleft}
\hypersetup{urlcolor=black}
{\Large The ATLAS Collaboration}

\bigskip

\AtlasOrcid[0000-0002-6665-4934]{G.~Aad}$^\textrm{\scriptsize 102}$,
\AtlasOrcid[0000-0002-5888-2734]{B.~Abbott}$^\textrm{\scriptsize 120}$,
\AtlasOrcid[0000-0002-1002-1652]{K.~Abeling}$^\textrm{\scriptsize 55}$,
\AtlasOrcid[0000-0001-5763-2760]{N.J.~Abicht}$^\textrm{\scriptsize 49}$,
\AtlasOrcid[0000-0002-8496-9294]{S.H.~Abidi}$^\textrm{\scriptsize 29}$,
\AtlasOrcid[0000-0002-9987-2292]{A.~Aboulhorma}$^\textrm{\scriptsize 35e}$,
\AtlasOrcid[0000-0001-5329-6640]{H.~Abramowicz}$^\textrm{\scriptsize 151}$,
\AtlasOrcid[0000-0002-1599-2896]{H.~Abreu}$^\textrm{\scriptsize 150}$,
\AtlasOrcid[0000-0003-0403-3697]{Y.~Abulaiti}$^\textrm{\scriptsize 117}$,
\AtlasOrcid[0000-0002-8588-9157]{B.S.~Acharya}$^\textrm{\scriptsize 69a,69b,m}$
% ...
\bigskip
\\

% ...
$^{29}$Physics Department, Brookhaven National Laboratory, Upton NY; United States of America.\\
% ...
$^{35}$$^{(a)}$Facult\'e des Sciences Ain Chock, R\'eseau Universitaire de Physique des Hautes Energies - Universit\'e Hassan II, Casablanca;$^{(b)}$Facult\'{e} des Sciences, Universit\'{e} Ibn-Tofail, K\'{e}nitra;$^{(c)}$Facult\'e des Sciences Semlalia, Universit\'e Cadi Ayyad, LPHEA-Marrakech;$^{(d)}$LPMR, Facult\'e des Sciences, Universit\'e Mohamed Premier, Oujda;$^{(e)}$Facult\'e des sciences, Universit\'e Mohammed V, Rabat;$^{(f)}$Institute of Applied Physics, Mohammed VI Polytechnic University, Ben Guerir; Morocco.\\
% ...
$^{49}$Fakult\"{a}t Physik , Technische Universit{\"a}t Dortmund, Dortmund; Germany.\\
% ...
$^{55}$II. Physikalisches Institut, Georg-August-Universit\"{a}t G\"ottingen, G\"ottingen; Germany.\\
% ...
$^{69}$$^{(a)}$INFN Gruppo Collegato di Udine, Sezione di Trieste, Udine;$^{(b)}$ICTP, Trieste;$^{(c)}$Dipartimento Politecnico di Ingegneria e Architettura, Universit\`a di Udine, Udine; Italy.\\
% ...
$^{102}$CPPM, Aix-Marseille Universit\'e, CNRS/IN2P3, Marseille; France.\\
% ...
$^{117}$Department of Physics, New York University, New York NY; United States of America.\\
% ...
$^{120}$Homer L. Dodge Department of Physics and Astronomy, University of Oklahoma, Norman OK; United States of America.\\
% ...
$^{150}$Department of Physics, Technion, Israel Institute of Technology, Haifa; Israel.\\
$^{151}$Raymond and Beverly Sackler School of Physics and Astronomy, Tel Aviv University, Tel Aviv; Israel.\\
% ...
$^{a}$ Also Affiliated with an institute covered by a cooperation agreement with CERN.\\
$^{b}$ Also at An-Najah National University, Nablus; Palestine.\\
$^{c}$ Also at Borough of Manhattan Community College, City University of New York, New York NY; United States of America.\\
$^{d}$ Also at Center for High Energy Physics, Peking University; China.\\
$^{e}$ Also at Center for Interdisciplinary Research and Innovation (CIRI-AUTH), Thessaloniki; Greece.\\
$^{f}$ Also at Centro Studi e Ricerche Enrico Fermi; Italy.\\
% ...
$^{m}$ Also at Department of Physics, King's College London, London; United Kingdom.\\
% ...

\end{flushleft}


%% file: ANA-EXOT-2019-33-PAPER.bib
@article{Dirac1931,
      author         = "Dirac, P.A.M. ",
      title          = "{Quantised Singularities in the Electromagnetic Field}",
      journal        = "Proc. Roy. Soc.",
      volume         = "A 133",
      pages          = "60",
      year           = "1931",
}

@article{Dirac1948,
      author         = "Dirac, P.A.M.",
      title          = "{The Theory of Magnetic Poles}",
      journal        = "Phys. Rev.",
      volume         = "74",
      pages          = "817",
      doi            = "10.1103/PhysRev.74.817",
      year           = "1948",
}

@article{Polyakov1974,
      author         = "Polyakov, A.M.",
      title          = "{Particle Spectrum in the Quantum Field Theory}",
      journal        = "JETP Lett.",
      volume         = "20",
      pages          = "194",
      year           = "1974",
}

@article{tHooft1974,
      author         = "'t Hooft, G.",
      title          = "{Magnetic monopoles in unified gauge theories}",
      journal        = "Nucl. Phys. B",
      volume         = "79",
      pages          = "276",
      doi            = "10.1016/0550-3213(74)90486-6",
      year           = "1974",
}

@article{Kephart2017,
    author = "Kephart, Thomas W. and Leontaris, George K. and Shafi, Qaisar",
    title = "{Magnetic monopoles and free fractionally charged states at accelerators and in cosmic rays}",
    doi = "10.1007/JHEP10(2017)176",
    journal = "JHEP",
    volume = "10",
    pages = "176",
    year = "2017",
    eprint = "1707.08067",
    archivePrefix = "arXiv",
    primaryClass = "hep-ph",
}

@article{CHO1997360,
  author = "Y. M. Cho and D. Maison",
  title = "Monopole configuration in {W}einberg-{S}alam model",
  journal = "Phys. Lett. B",
  volume = "391",
  pages = "360",
  year = "1997",
  doi = "10.1016/S0370-2693(96)01492-X",
  archivePrefix = "arXiv",
  eprint = "hep-th/9601028",
}

@article{Kimm_2015,
   author="K. Kimm and J. H. Yoon and Y. M. Cho",
   title="Finite energy electroweak dyon",
   volume="75",
   journal="Eur. Phys. J. C",
   year="2015",
   Pages = "67",
   doi = "10.1140/epjc/s10052-015-3290-3",
   archivePrefix = "arXiv",
   eprint = "1305.1699",
   primaryClass = "hep-ph",
}

@article{Ellis2016,
      author        = "Ellis, J. and Mavromatos, N. E. and You, T.",
      title         = "The price of an electroweak monopole",
      journal       = "Phys. Lett. B",
      volume        = "756",
      pages         = "29",
      year          = "2016",
      doi           = "10.1016/j.physletb.2016.02.048",
      archivePrefix = "arXiv",
      eprint        = "1602.01745",
      primaryClass  = "hep-ph",
}

@article{Arunasalam2017,
        author = "Arunasalam, S. and Kobakhidze, A.",
        title = "Electroweak monopoles and the electroweak phase transition",
        journal = "Eur. Phys. J. C",
        year = 2017,
        volume = 77,
        pages = "444",
        doi   = "10.1140/epjc/s10052-017-4999-y",
        archivePrefix = "arXiv",
        eprint = "1702.04068",
        primaryClass = "hep-ph",
}

@article{Ellis2017,
  author = "Ellis, J. and Mavromatos, N. E. and You, T.",
  title = "{Light-by-Light Scattering Constraint on Born-Infeld Theory}",
  journal = "{Phys. Rev. Lett.}",
  volume = "118",
  pages = "261802",
  year = "2017",
  doi = "10.1103/PhysRevLett.118.261802",
  archivePrefix = "arXiv",
  eprint = "1703.08450",
  primaryClass = "hep-ph",
}

@article{Arai2018,
   author = "Arai, Masato and Blaschke, Filip and Eto, Minoru and Sakai, Norisuke",
   title = "{Localization of the Standard Model via the Higgs mechanism and a finite electroweak monopole from non-compact five dimensions}",
   volume = "2018",
   DOI = "10.1093/ptep/pty083",
   number = "8",
   journal = "PTEP",
   year = "2018",
   archivePrefix = "arXiv",
   eprint = "1802.06649",
   primaryClass = "hep-ph",
}

@article{Mavromatos2017,
   author = "Mavromatos, Nick E. and Sarkar, Sarben",
   title = "{Magnetic monopoles from global monopoles in the presence of a Kalb-Ramond Field}",
   volume = "95",
   DOI = "10.1103/physrevd.95.104025",
   number = "10",
   pages = "104025",
   journal = "Phys. Rev. D",
   year = "2017",
   archivePrefix = "arXiv",
   eprint = "1607.01315",
   primaryClass = "hep-th",
}

@article{Mavromatos2018,
   author = "Mavromatos, Nick E. and Sarkar, Sarben",
   title = "{Regularized Kalb-Ramond magnetic monopole with finite energy}",
   volume = "97",
   DOI = "10.1103/physrevd.97.125010",
   number = "12",
   pages = "125010",
   journal = "Phys. Rev. D",
   year = "2018",
   archivePrefix = "arXiv",
   eprint = "1804.01702",
   primaryClass = "hep-th",
}

@article{Cho2019,
  title = "{Physical implications of electroweak monopole}",
  author = "Cho, Y. M.",
  journal = "Phil. Trans. Roy. Soc. Lond. A",
  year = "2019",
  volume = "377",
  pages = "20190038",
  doi = "10.1098/rsta.2019.0038"
}

@article{Hung2020,
    author = "Hung, P. Q.",
    title = "{Topologically stable, finite-energy electroweak-scale monopoles}",
    eprint = "2003.02794",
    archivePrefix = "arXiv",
    primaryClass = "hep-ph",
    doi = "10.1016/j.nuclphysb.2020.115278",
    journal = "Nucl. Phys. B",
    volume = "962",
    pages = "115278",
    year = "2021"
}

@article{Ellis2020,
    author = "Ellis, John and Hung, P. Q. and Mavromatos, Nick E.",
    title = "{An electroweak monopole, Dirac quantization and the weak mixing angle}",
    eprint = "2008.00464",
    archivePrefix = "arXiv",
    primaryClass = "hep-ph",
    doi = "10.1016/j.nuclphysb.2021.115468",
    journal = "Nucl. Phys. B",
    volume = "969",
    pages = "115468",
    year = "2021"
}

@article{Kibble1976,
    author = "Kibble, T. W. B.",
    title = "{Topology of cosmic domains and strings}",
    reportNumber = "ICTP/75/5",
    doi = "10.1088/0305-4470/9/8/029",
    journal = "J. Phys. A",
    volume = "9",
    pages = "1387",
    year = "1976"
}

@article{Chin1979,
  title = "Possible Long-Lived Hyperstrange Multiquark Droplets",
  author = "Chin, S. A. and Kerman, A. K.",
  journal = "Phys. Rev. Lett.",
  volume = "43",
  pages = "1292",
  year = "1979",
  doi = "10.1103/PhysRevLett.43.1292"
}

@article{Farhi1984,
      author         = "Farhi, E. and Jaffe, R.L.",
      title          = "{Strange matter}",
      journal        = "Phys. Rev. D",
      volume         = "30",
      pages          = "2379",
      doi            = "10.1103/PhysRevD.30.2379",
      year           = "1984",
}

@article{Coleman1985,
      author         = "Coleman, S.R.",
      title          = "{Q-Balls}",
      journal        = "Nucl. Phys. B",
      volume         = "262",
      pages          = "263",
      doi            = "10.1016/0550-3213(85)90286-X",
      year           = "1985",
}

@article{Kusenko1997,
title = "Solitons in the supersymmetric extensions of the standard model",
journal = "Phys. Lett. B",
volume = "405",
number = "1",
pages = "108",
year = "1997",
doi = "10.1016/S0370-2693(97)00584-4",
author = "Alexander Kusenko",
}

@article{udQM2018,
  title = {Quark Matter May Not Be Strange},
  author = {Holdom, Bob and Ren, Jing and Zhang, Chen},
  journal = {Phys. Rev. Lett.},
  volume = {120},
  issue = {22},
  pages = {222001},
  numpages = {6},
  year = {2018},
  publisher = {American Physical Society},
  doi = {10.1103/PhysRevLett.120.222001},
  eprint = "1707.06610",
  archivePrefix = "arXiv",
  primaryClass = "hep-ph",
}

@article{DeRujula1984,
  title = "{Nuclearites—a novel form of cosmic radiation}",
  author = "{De R{\'u}jula, A. and Glashow, S. L.}",
  journal = "Nature",
  volume = "312",
  pages = "734",
  year = "1984",
  doi = "10.1038/312734a0",
}

@article{Giacomelli2003,
    author = "Giacomelli, G. and Patrizii, L.",
    title = "Magnetic monopole searches",
    journal = "ICTP Lect. Notes Ser.",
    volume = "14",
    pages = "121",
    year = "2003",
    archivePrefix = "arXiv",
    eprint = "hep-ex/0302011",
}

@article{Milton2006,
    author = {Kimball A Milton},
    title = {Theoretical and experimental status of magnetic monopoles},
    journal = {Reports on Progress in Physics},
    doi = {10.1088/0034-4885/69/6/R02},
    volume = {69},
    number = {6},
    pages = {1637},
    year = {2006},
}

@article{Burdin2015,
      author         = "Burdin, S. and Fairbairn, M. and Mermod, P. and Milstead, D. and Pinfold, J. and  Sloan, T. and Taylor, W.",
      title          = "{Non-collider searches for stable massive particles}",
      journal        = "Phys. Rept.",
      volume         = "582",
      year           = "2015",
      pages          = "1",
      doi            = "10.1016/j.physrep.2015.03.004",
      eprint         = "1410.1374",
      archivePrefix  = "arXiv",
      primaryClass   = "hep-ph",
}

@article{Patrizii2015,
      author         = "Patrizii, L. and Spurio, M.",
      title          = "{Status of Searches for Magnetic Monopoles}",
      journal        = "Ann. Rev. Nucl. Part. Sci.",
      volume         = "65",
      year           = "2015",
      pages          = "279",
      doi            = "10.1146/annurev-nucl-102014-022137",
      eprint         = "1510.07125",
      archivePrefix  = "arXiv",
      primaryClass   = "hep-ex"
}

@article{Patrizii2019,
    author = "Patrizii, L. and Sahnoun, Z. and Togo, V.",
    title = "{Searches for cosmic magnetic monopoles: past, present and future}",
    doi = "10.1098/rsta.2018.0328",
    journal = "Phil. Trans. Roy. Soc. Lond. A",
    volume = "377",
    number = "2161",
    pages = "20180328",
    year = "2019"
}

@article{Mavromatos2020,
   author = "Mavromatos, Nick E. and Mitsou, Vasiliki A.",
   title = "Magnetic monopoles revisited: Models and searches at colliders and in the Cosmos",
   journal = "Int. J. Mod. Phys. A",
   volume = "35",
   number = "23",
   pages = "2030012",
   year = "2020",
   doi = "10.1142/s0217751x20300124",
   archivePrefix = "arXiv",
   eprint = "2005.05100",
   primaryClass = "hep-ph",
}

@article{Battistoni:1983,
    author = "Battistoni, G. and others",
    title = "{Nucleon Stability, Magnetic Monopoles and Atmospheric Neutrinos in the Mont Blanc Experiment}",
    doi = "10.1016/0370-2693(83)90827-4",
    journal = "Phys. Lett. B",
    volume = "133",
    pages = "454",
    year = "1983"
}

@article{Krishnaswamy:1984,
    author = "Krishnaswamy, M. R. and Menon, M. G. K. and Mondal, N. K. and Narasimham, V. S. and Sreekantan, B. V. and Hayashi, Y. and Ito, N. and Kawakami, S. and Miyake, S.",
    title = "{Limits on the flux of monopoles from the Kolar Gold Mine experiments}",
    doi = "10.1016/0370-2693(84)91143-2",
    journal = "Phys. Lett. B",
    volume = "142",
    pages = "99",
    year = "1984"
}

@article{Orito:1991,
    author = "Orito, S. and others",
    title = "{Search for supermassive relics with a 2000-m$^2$ array of plastic track detectors}",
    doi = "10.1103/PhysRevLett.66.1951",
    journal = "Phys. Rev. Lett.",
    volume = "66",
    pages = "1951",
    year = "1991"
}

@article{Soudan-2:1992,
    %author = "Soudan-2 Collaboration",
    author = "Thron, J. L. and others",
    title = "{Search for magnetic monopoles with the Soudan 2 detector}",
    doi = "10.1103/PhysRevD.46.4846",
    journal = "Phys. Rev. D",
    volume = "46",
    pages = "4846",
    year = "1992"
}

@article{MACRO:1994,
    %author = "MACRO Collaboration",
    author = "Ahlen, S. and others", 
    title = "{Search for slowly moving magnetic monopoles with the MACRO detector}",
    doi = "10.1103/PhysRevLett.72.608",
    journal = "Phys. Rev. Lett.",
    volume = "72",
    pages = "608",
    year = "1994"
}

@article{MACRO:1997,
    %author = "MACRO Collaboration",
    author = "Ambrosio, M. and others",
    title = "{Magnetic monopole search with the MACRO detector at Gran Sasso}",
    doi = "10.1016/S0370-2693(97)00684-9",
    journal = "Phys. Lett. B",
    volume = "406",
    pages = "249",
    year = "1997"
}

@article{MACRO:2000,
    %author = "MACRO Collaboration",
    author = "Ambrosio, M. and others",
    title = "{Nuclearite search with the macro detector at Gran Sasso}",
    eprint = "hep-ex/9904031",
    archivePrefix = "arXiv",
    doi = "10.1007/s100520050708",
    journal = "Eur. Phys. J. C",
    volume = "13",
    pages = "453",
    year = "2000"
}

@article{MACRO:2002a,
    %author = "MACRO Collaboration",
    author = "Ambrosio, M. and others",
    title = "{Final results of magnetic monopole searches with the MACRO experiment}",
    eprint = "hep-ex/0207020",
    archivePrefix = "arXiv",
    reportNumber = "DFUB-02-05",
    doi = "10.1140/epjc/s2002-01046-9",
    journal = "Eur. Phys. J. C",
    volume = "25",
    pages = "511",
    year = "2002"
}

@article{Baksan:2006,
    author = "Novoseltsev, Yu. F. and Boliev, M. M. and Butkevich, A. V. and Mikheev, S. P. and Petkov, V. B.",
    title = "{A search for massive magnetic monopoles at the Baksan Underground Scintillation Telescope (BUST)}",
    doi = "10.1016/j.nuclphysbps.2005.07.048",
    journal = "Nucl. Phys. B Proc. Suppl.",
    volume = "151",
    pages = "337",
    year = "2006"
}

@article{BAIKAL:2008,
    %author = "BAIKAL Collaboration",
    author = "Aynutdinov, V. and others",
    title = "{Search for relativistic magnetic monopoles with the Baikal neutrino telescope}",
    doi = "10.1016/j.astropartphys.2008.03.006",
    journal = "Astropart. Phys.",
    volume = "29",
    pages = "366",
    year = "2008"
}

@article{SLIM:2008a,
    author = "Balestra, S. and others",
    title = "{Magnetic Monopole Search at high altitude with the SLIM experiment}",
    eprint = "0801.4913",
    archivePrefix = "arXiv",
    primaryClass = "hep-ex",
    doi = "10.1140/epjc/s10052-008-0597-3",
    journal = "Eur. Phys. J. C",
    volume = "55",
    pages = "57",
    year = "2008"
}

@article{SLIM:2008b,
    %author = "SLIM Collaboration",
    author = "Cecchini, S. and others",
    title = "{Results of the search for strange quark matter and Q-balls with the SLIM experiment}",
    eprint = "0805.1797",
    archivePrefix = "arXiv",
    primaryClass = "hep-ex",
    doi = "10.1140/epjc/s10052-008-0747-7",
    journal = "Eur. Phys. J. C",
    volume = "57",
    pages = "525",
    year = "2008"
}

@article{RICE:2008,
    author = "Hogan, D. P. and Besson, D. Z. and Ralston, J. P. and Kravchenko, I. and Seckel, D.",
    title = "{Relativistic Magnetic Monopole Flux Constraints from RICE}",
    eprint = "0806.2129",
    archivePrefix = "arXiv",
    primaryClass = "astro-ph",
    doi = "10.1103/PhysRevD.78.075031",
    journal = "Phys. Rev. D",
    volume = "78",
    pages = "075031",
    year = "2008"
}

@article{AMANDA:2010,
    author = "Abbasi, R. and others",
    title = "{Search for relativistic magnetic monopoles with the AMANDA-II neutrino telescope}",
    doi = "10.1140/epjc/s10052-010-1411-6",
    journal = "Eur. Phys. J. C",
    volume = "69",
    pages = "361",
    year = "2010"
}

@article{ANITA-II:2010,
    author = {Detrixhe, M. and Besson, D. and Gorham, P. W. and Allison, P. and Baughmann, B. and Beatty, J. J. and Belov, K. and Bevan, S. and Binns, W. R. and Chen, C. and Chen, P. and Clem, J. M. and Connolly, A. and De Marco, D. and Dowkontt, P. F. and DuVernois, M. A. and Frankenfeld, C. and Grashorn, E. W. and Hogan, D. P. and Griffith, N. and Hill, B. and Hoover, S. and Israel, M. H. and Javaid, A. and Liewer, K. M. and Matsuno, S. and Mercurio, B. C. and Miki, C. and Mottram, M. and Nam, J. and Nichol, R. J. and Palladino, K. and Romero-Wolf, A. and Ruckman, L. and Saltzberg, D. and Seckel, D. and Varner, G. S. and Vieregg, A. G. and Wang, Y.},
    collaboration = {ANITA Collaboration},
    title = "{Ultrarelativistic magnetic monopole search with the ANITA-II balloon-borne radio interferometer}",
    eprint = "1008.1282",
    archivePrefix = "arXiv",
    primaryClass = "astro-ph.HE",
    doi = "10.1103/PhysRevD.83.023513",
    journal = "Phys. Rev. D",
    volume = "83",
    pages = "023513",
    year = "2011"
}

@article{PierreAuger:2016,
    collaboration = "{Pierre Auger Collaboration}",
    author = "Aab, A. and others",
    title = "{Search for ultrarelativistic magnetic monopoles with the Pierre Auger observatory}",
    eprint = "1609.04451",
    archivePrefix = "arXiv",
    primaryClass = "astro-ph.HE",
    doi = "10.1103/PhysRevD.94.082002",
    journal = "Phys. Rev. D",
    volume = "94",
    number = "8",
    pages = "082002",
    year = "2016"
}

@article{ANTARES:2012,
    collaboration = "ANTARES Collaboration",
    author = "Adrián-Martínez, S. and others",
    title = "{Search for relativistic magnetic monopoles with the ANTARES neutrino telescope}",
    eprint = "1110.2656",
    archivePrefix = "arXiv",
    primaryClass = "astro-ph.HE",
    doi = "10.1016/j.astropartphys.2012.02.007",
    journal = "Astropart. Phys.",
    volume = "35",
    pages = "634",
    year = "2012"
}

@article{ANTARES:2017,
    author = "Albert, A. and others",
    collaboration = "ANTARES Collaboration",
    title = "{Search for relativistic magnetic monopoles with five years of the ANTARES detector data}",
    eprint = "1703.00424",
    archivePrefix = "arXiv",
    primaryClass = "astro-ph.HE",
    doi = "10.1007/JHEP07(2017)054",
    journal = "JHEP",
    volume = "07",
    pages = "054",
    year = "2017"
}

@article{ANTARES:2021b,
    %author = "ANTARES Collaboration",
    author = "Bouta, M. and others", 
    title = "{Nuclearite search with ANTARES}",
    eprint = "2107.13479",
    archivePrefix = "arXiv",
    primaryClass = "astro-ph.HE",
    doi = "10.1088/1748-0221/16/09/C09010",
    journal = "JINST",
    volume = "16",
    number = "09",
    pages = "C09010",
    year = "2021"
}

@article{ANTARES:2021a,
    author = "Boumaaza, J. and others",
    %author = "ANTARES Collaboration",
    title = "{Search for relativistic magnetic monopoles with ten years of the ANTARES detector data}",
    doi = "10.1088/1748-0221/16/11/C11004",
    journal = "JINST",
    volume = "16",
    number = "11",
    pages = "C11004",
    year = "2021"
}

@article{ANTARES:2022,
    %author = "ANTARES Collaboration",
    author = "Albert, A. and others",
    title = "{Search for magnetic monopoles with ten years of the ANTARES neutrino telescope}",
    eprint = "2202.13786",
    archivePrefix = "arXiv",
    primaryClass = "astro-ph.HE",
    doi = "10.1016/j.jheap.2022.03.001",
    journal = "JHEAp",
    volume = "34",
    pages = "1",
    year = "2022"
}

@article{ANTARES:2023,
    %author = "ANTARES Collaboration",
    author = "Albert, A. and others",
    title = "{Limits on the nuclearite flux using the ANTARES neutrino telescope}",
    eprint = "2208.11689",
    archivePrefix = "arXiv",
    primaryClass = "astro-ph.HE",
    doi = "10.1088/1475-7516/2023/01/012",
    journal = "JCAP",
    volume = "01",
    pages = "012",
    year = "2023"
}

@article{IceCube:2013,
    collaboration = "IceCube Collaboration",
    author = "Abbasi, R. and others",
    title = "{Search for relativistic magnetic monopoles with IceCube}",
    eprint = "1208.4861",
    archivePrefix = "arXiv",
    primaryClass = "astro-ph.HE",
    doi = "10.1103/PhysRevD.87.022001",
    journal = "Phys. Rev. D",
    volume = "87",
    number = "2",
    pages = "022001",
    year = "2013"
}

@article{IceCube:2016a,
    author = "{A. Obertacke for the IceCube Collaboration}",
    %author = "Obertacke, Anna",
    title = "{New limit for mildly relativistic magnetic monopoles obtained with IceCube}",
    doi = "10.22323/1.236.1061",
    journal = "PoS",
    volume = "ICRC2015",
    pages = "1061",
    year = "2016"
}

@article{IceCube:2016b,
    %author = "IceCube Collaboration",
    author = "Aartsen, M.G. and others",
    title = "{Searches for relativistic magnetic monopoles in IceCube}",
    eprint = "1511.01350",
    archivePrefix = "arXiv",
    primaryClass = "astro-ph.HE",
    doi = "10.1140/epjc/s10052-016-3953-8",
    journal = "Eur. Phys. J. C",
    volume = "76",
    number = "3",
    pages = "133",
    year = "2016"
}

@article{IceCube:2018,
    collaboration = "IceCube Collaboration",
    author = "{A. Pollmann for the IceCube Collaboration}",  
    title = "{Searches for magnetic monopoles with IceCube}",
    doi = "10.1051/epjconf/201816804010",
    journal = "EPJ Web Conf.",
    volume = "168",
    pages = "04010",
    year = "2018"
}

@article{IceCube:2022,
    author = "Abbasi, R. and others",
    collaboration = "IceCube Collaboration",
    title = "{Search for Relativistic Magnetic Monopoles with Eight Years of IceCube Data}",
    eprint = "2109.13719",
    archivePrefix = "arXiv",
    primaryClass = "astro-ph.HE",
    doi = "10.1103/PhysRevLett.128.051101",
    journal = "Phys. Rev. Lett.",
    volume = "128",
    number = "5",
    pages = "051101",
    year = "2022"
}

@phdthesis{NOvA:2015,
    author = "Wang, Z.",
    title = "{Search for Magnetic Monopoles with the NOvA Far Detector}",
    reportNumber = "FERMILAB-THESIS-2015-28",
    doi = "10.18130/V3FR9B",
    school = "Physics - Graduate School of Arts and Sciences, University of Virginia, Virginia U.",
    year = "2015"
}

@article{NOvA:2021,
    author = "Acero, M. A. and others",
    collaboration = "NOvA Collaboration",
    title = "{Search for slow magnetic monopoles with the NOvA detector on the surface}",
    eprint = "2009.04867",
    archivePrefix = "arXiv",
    primaryClass = "hep-ex",
    doi = "10.1103/PhysRevD.103.012007",
    journal = "Phys. Rev. D",
    volume = "103",
    number = "1",
    pages = "012007",
    year = "2021"
}

@article{PETRA:1983,
    author = "Musset, P. and Price, M. and Lohrmann, E.",
    title = "{Search for Magnetic Monopoles in Electron - Positron Collisions at 34-{GeV} Center-of-mass Energy}",
    reportNumber = "PRINT-83-0602 (HAMBURG), DESY-83-053",
    doi = "10.1016/0370-2693(83)90270-8",
    journal = "Phys. Lett. B",
    volume = "128",
    pages = "333",
    year = "1983"
}

@article{PEP:1982,
    author = "Kinoshita, K. and Price, P. B. and Fryberger, D.",
    title = "{Search for Highly Ionizing Particles in ${e}^{+}{e}^{\ensuremath{-}}$ Collisions at $\sqrt{s}=29$ GeV}",
    doi = "10.1103/PhysRevLett.48.77",
    journal = "Phys. Rev. Lett.",
    volume = "48",
    pages = "77",
    year = "1982"
}

@article{PEP:1984,
    author = "Fryberger, D. and Coan, T. E. and Kinoshita, K. and Price, P. B.",
    title = "{Search for Highly Ionizing Particles in $e^+ e^-$ Collisions at $\sqrt{s}=29$ {GeV}}",
    doi = "10.1103/PhysRevD.29.1524",
    journal = "Phys. Rev. D",
    volume = "29",
    pages = "1524",
    year = "1984"
}

@article{CLEO:1987,
    author = "Gentile, T. and others", 
    %collaboration = "CLEO Collaboration",
    title = "{Search for magnetically charged particles produced in $e^+ e^-$ annihilations at $\sqrt{s}=10.6$~GeV}",
    doi = "10.1103/PhysRevD.35.1081",
    journal = "Phys. Rev. D",
    volume = "35",
    pages = "1081",
    year = "1987"
}

@article{TASSO:1988,
    author = "Braunschweig, W. and others",
    collaboration = "TASSO",
    title = "{A search for particles With magnetic charge produced in $e^+ e^-$ annihilations at $\sqrt{s}=35$ GeV}",
    doi = "10.1007/BF01624358",
    journal = "Z. Phys. C",
    volume = "38",
    pages = "543",
    year = "1988"
}

@article{TRISTAN:1988,
    author = "Kinoshita, K. and Fujii, M. and Nakajima, K. and Price, P. B. and Tasaka, S.",
    title = "{Search for highly ionizing particles in $e^+ e^-$ annihilations at $\sqrt{s}=50$-$52$ GeV}",
    doi = "10.1103/PhysRevLett.60.1610",
    journal = "Phys. Rev. Lett.",
    volume = "60",
    pages = "1610",
    year = "1988"
}

@article{TRISTAN:1989,
    author = "Kinoshita, K. and Fujii, M. and Nakajima, K. and Price, P. B. and Tasaka, S.",
    title = "{Search for highly ionizing particles in $e^+ e^-$ annihilations at $\sqrt{s}=50$-$60.8$ GeV}",
    doi = "10.1016/0370-2693(89)90990-8",
    journal = "Phys. Lett. B",
    volume = "228",
    pages = "543",
    year = "1989"
}

@article{OPAL:2008,
    author = "G. Abbiendi and others", 
    collaboration = "OPAL Collaboration",
    title = "{Search for Dirac magnetic monopoles in $e^+e^-$ collisions with the OPAL detector at LEP2}",
    eprint = "0707.0404",
    archivePrefix = "arXiv",
    primaryClass = "hep-ex",
    doi = "10.1016/j.physletb.2008.03.057",
    journal = "Phys. Lett. B",
    volume = "663",
    pages = "37",
    year = "2008"
}

@article{MODAL:1992,
    author = "Kinoshita, K. and Du, R. and Giacomelli, G. and Patrizii, L. and Predieri, F. and Serra, P. and Spurio, M. and Pinfold, J. L.",
    title = "{Search for highly ionizing particles in $e^+ e^-$ annihilations at $\sqrt{s}=91.1$ GeV}",
    doi = "10.1103/PhysRevD.46.R881",
    journal = "Phys. Rev. D",
    volume = "46",
    pages = "R881",
    year = "1992"
}

@article{L6-MODAL:1993,
    author = "Pinfold, J. L. and Du, R. and Kinoshita, K. and Lorazo, B. and Regimbald, M. and Price, B.",
    title = "{A Search for highly ionizing particles produced at the OPAL intersection point at LEP}",
    doi = "10.1016/0370-2693(93)90346-J",
    journal = "Phys. Lett. B",
    volume = "316",
    pages = "407",
    year = "1993"
}

@article{H1:2004,
    author = "{H1 Collaboration}",
    collaboration = "H1 Collaboration",
    title = "{A direct search for stable magnetic monopoles produced in positron-proton collisions at HERA}",
    eprint = "hep-ex/0501039",
    reportNumber = "DESY-04-240",
    doi = "10.1140/epjc/s2005-02201-6",
    journal = "Eur. Phys. J. C",
    volume = "41",
    pages = "133",
    year = "2005"
}

@article{Johnsen:1973fop,
    author = "Johnsen, K.",
    title = "{CERN Intersecting Storage Rings (ISR)}",
    doi = "10.1073/pnas.70.2.619",
    journal = "Proc. Nat. Acad. Sci.",
    volume = "70",
    pages = "619",
    year = "1973"
}

@article{CDF:2005,
    author = "Abulencia, A. and others",
    collaboration = "CDF Collaboration",
    title = "{Direct Search for Dirac Magnetic Monopoles in $p\bar{p}$ Collisions at $\sqrt{s} = 1.96$ TeV}",
    eprint = "hep-ex/0509015",
    archivePrefix = "arXiv",
    doi = "10.1103/PhysRevLett.96.201801",
    journal = "Phys. Rev. Lett.",
    volume = "96",
    pages = "201801",
    year = "2006"
}

@article{Kalbfleisch:2003,
    author = "Kalbfleisch, G. R. and Luo, W. and Milton, K. A. and Smith, E. H. and Strauss, Michael G.",
    title = "{Limits on production of magnetic monopoles utilizing samples from the D0 and CDF detectors at the Fermilab Tevatron}",
    eprint = "hep-ex/0306045",
    archivePrefix = "arXiv",
    doi = "10.1103/PhysRevD.69.052002",
    journal = "Phys. Rev. D",
    volume = "69",
    pages = "052002",
    year = "2004"
}

@article{Kalbfleisch:2000,
    author = "Kalbfleisch, G. R. and Milton, K. A. and Strauss, Michael G. and Gamberg, L. P. and Smith, E. H. and Luo, W.",
    title = "{Improved Experimental Limits on the Production of Magnetic Monopoles}",
    eprint = "hep-ex/0005005",
    archivePrefix = "arXiv",
    doi = "10.1103/PhysRevLett.85.5292",
    journal = "Phys. Rev. Lett.",
    volume = "85",
    pages = "5292",
    year = "2000"
}

@article{D0:1998,
  author = "Abbott, B. and others",
  title = {Search for Heavy Pointlike Dirac Monopoles},
  collaboration = {D0 Collaboration},
  journal = {Phys. Rev. Lett.},
  volume = {81},
  pages = {524},
  year = {1998},
  doi = {10.1103/PhysRevLett.81.524},
  eprint = "hep-ex/9803023",
  archivePrefix = "arXiv",
}

@article{Price:1987,
  title = {Search for Highly Ionizing Particles at the Fermilab Proton-Antiproton Collider},
  author = {Price, P. B. and Ren, G. and Kinoshita, K.},
  journal = {Phys. Rev. Lett.},
  volume = {59},
  pages = {2523},
  year = {1987},
  doi = "10.1103/PhysRevLett.59.2523"
}

@article{Price:1990,
    author = "Price, P. B. and Jing, G. and Kinoshita, K.",
    title = "{High luminosity search for highly ionizing particles at the Fermilab collider}",
    doi = "10.1103/PhysRevLett.65.149",
    journal = "Phys. Rev. Lett.",
    volume = "65",
    pages = "149",
    year = "1990"
}

@article{Bertani:1990,
    author = "Bertani, M. and others",
    title = "{Search for Magnetic Monopoles at the Tevatron Collider}",
    doi = "10.1209/0295-5075/12/7/007",
    journal = "EPL",
    volume = "12",
    pages = "613",
    year = "1990"
}

@article{Hoffmann:1978,
    author = "Hoffmann, H. and Kantardjian, G. and Di Liberto, S. and Meddi, F. and Romano, G. and Rosa, G.",
    title = "{A new search for magnetic monopoles at the CERN ISR with plastic detectors}",
    doi = "10.1007/BF02776247",
    journal = "Lett. Nuovo Cim.",
    volume = "23",
    pages = "357",
    year = "1978"
}

@article{Carrigan:1977,
    author = "Carrigan, Jr., Richard A. and Strauss, B. P. and Giacomelli, G.",
    title = "{Search for magnetic monopoles at the CERN Intersecting Storage Rings}",
    doi = "10.1103/PhysRevD.17.1754",
    journal = "Phys. Rev. D",
    volume = "17",
    pages = "1754",
    year = "1978"
}

@article{Aubert:1982,
    author = "Aubert, B. and Musset, P. and Price, M. and Vialle, J. P.",
    title = "{Search for magnetic monopoles in proton-Antiproton interactions at 540~GeV cm energy}",
    doi = "10.1016/0370-2693(83)90488-4",
    journal = "Phys. Lett. B",
    volume = "120",
    pages = "465",
    year = "1983"
}

@article{Carrigan:1973,
    author = "Carrigan, R. A. and Nezrick, F. A. and Strauss, B. P.",
    title = "{Search for Magnetic Monopole Production by 300-GeV Protons}",
    doi = "10.1103/PhysRevD.8.3717",
    journal = "Phys. Rev. D",
    volume = "8",
    pages = "3717",
    year = "1973"
}

@article{Carrigan:1974,
  title = "{Extension of Fermi National Accelerator Laboratory magnetic-monopole search to 400 GeV}",
  author = "Carrigan, R. A. and Nezrick, F. A. and Strauss, B. P.",
  journal = "Phys. Rev. D",
  volume = "10",
  pages = "3867",
  year = "1974",
  doi = "10.1103/PhysRevD.10.3867"
}

@article{PhysRevLett.79.3134,
  author = "He, Y. D.",
  title = "{Search for a Dirac Magnetic Monopole in High Energy Nucleus-Nucleus Collisions}",
  journal = "Phys. Rev. Lett.",
  volume = "79",
  pages = "3134",
  year = "1997",
  doi = "10.1103/PhysRevLett.79.3134",
}

@article{STAR:2005,
    author = "Abelev, B. I. and others",
    collaboration = "STAR Collaboration",
    title = "{Strangelet search in Au+Au collisions at $\sqrt{s_{NN}}$ = 200 GeV}",
    eprint = "nucl-ex/0511047",
    archivePrefix = "arXiv",
    doi = "10.1103/PhysRevC.76.011901",
    journal = "Phys. Rev. C",
    volume = "76",
    pages = "011901",
    year = "2007"
}

@article{E886:1996,
    author = "Rusek, A. and others",
    collaboration = "E886 Collaboration",
    title = "{Strangelet search and light nucleus production in relativistic Si+Pt and Au+Pt collisions}",
    doi = "10.1103/PhysRevC.54.R15",
    journal = "Phys. Rev. C",
    volume = "54",
    pages = "R15",
    year = "1996"
}

@article{E864:1999,
    author = "{Gene Van Buren for the E864 Collaboration}",
    collaboration = "E864 Collaboration",
    title = "{Negatively charged strangelet search using the E864 spectrometer at the AGS}",
    eprint = "hep-ex/9811049",
    archivePrefix = "arXiv",
    doi = "10.1088/0954-3899/25/2/030",
    journal = "J. Phys. G",
    volume = "25",
    pages = "411",
    year = "1999"
}

@article{Xu_E864:1999,
author = "{Zhangbu Xu for the E864 Collaboration}",
title = {Search for positively charged strangelets and other related results with E864 at the AGS},
doi = {10.1088/0954-3899/25/2/029},
year = {1999},
volume = {25},
number = {2},
pages = {403},
journal = {J. Phys. G},
}

@article{ATLAS2020a,
      author         = "{ATLAS Collaboration}",
      title         = "{Search for heavy long-lived multi-charged particles in the full LHC Run 2 $pp$ collision data at $\sqrt{s}$ = 13 TeV using the ATLAS detector}",
      collaboration  = "ATLAS",
      year           = "2023",
      eprint         = "2303.13613",
      archivePrefix  = "arXiv",
      primaryClass   = "hep-ex",
      DOI            = "10.48550/arXiv.2303.13613",
}

@article{MoEDAL2016a,
      author         = "{MoEDAL Collaboration}",
      title          = "Search for magnetic monopoles with the {MoEDAL} prototype trapping detector in 8 {TeV} proton-proton collisions at the {LHC}",
      journal        = "JHEP",
      year           = "2016",
      volume         = "2016",
      number         = "08",
      pages          = "67",
      doi            = "10.1007/JHEP08(2016)067",
      archivePrefix  = "arXiv",
      eprint         = "1604.06645",
      primaryClass   = "hep-ex",
}

@article{MoEDAL2017a,
      author         = "{MoEDAL Collaboration}",
      title          = "{Search for Magnetic Monopoles with the MoEDAL Forward Trapping Detector in 13~TeV Proton-Proton Collisions at the LHC}",
      journal        = "Phys. Rev. Lett.",
      volume         = "118",
      year           = "2017",
      pages          = "061801",
      doi            = "10.1103/PhysRevLett.118.061801",
      archivePrefix  = "arXiv",
      eprint         = "1611.06817",
      primaryClass   = "hep-ex",
}

@article{MoEDAL2018,
   author = "{MoEDAL Collaboration}",
   title = "{Search for magnetic monopoles with the MoEDAL forward trapping
   detector in 2.11 fb$^{-1}$ of 13 TeV proton–proton collisions at the LHC}",
   volume = "782",
   doi = "10.1016/j.physletb.2018.05.069",
   journal = "Phys. Lett. B",
   year = "2018",
   archivePrefix  = "arXiv",
   eprint         = "1712.09849",
   primaryClass   = "hep-ex",
}

@article{MoEDAL2019,
  author  = "{MoEDAL Collaboration}",
  title   = "{Magnetic Monopole Search with the Full MoEDAL Trapping Detector in 13 TeV $pp$ Collisions Interpreted in Photon-Fusion and Drell-Yan Production}",
  journal = "Phys. Rev. Lett.",
  volume  = "123",
  pages   = "021802",
  year    = "2019",
  doi     = "10.1103/PhysRevLett.123.021802",
  archivePrefix = "arXiv",
  eprint  = "1903.08491",
  primaryClass = "hep-ex",
}

@article{dyonMoEDAL2021,
   author = "{MoEDAL Collaboration}",
   title = "First Search for Dyons with the Full MoEDAL Trapping Detector in 13 TeV $pp$ Collisions",
   volume = "126",
   doi = "10.1103/physrevlett.126.071801",
   number = "7",
   journal = "Phys. Rev. Lett.",
   year="2021",
   archivePrefix = "arXiv",
   eprint  = "2002.00861",
   primaryClass = "hep-ex",
}

@article{MoEDAL2021Schwinger,
      author         = "{MoEDAL Collaboration}",
      title          = "{Search for magnetic monopoles produced via the Schwinger mechanism}",
      archivePrefix  = "arXiv",
      eprint         = "2106.11933",
      primaryClass   = "hep-ex",
      doi = "10.1038/s41586-021-04298-1",
      journal = "Nature",
      volume = "602",
      number = "7895",
      pages = "63",
      year = "2022"
}

@article{MoEDAL2021HIP,
    author = "{MoEDAL Collaboration}",
    title = "{Search for highly-ionizing particles in $pp$ collisions at the LHC\textquoteright{}s Run-1 using the prototype MoEDAL detector}",
    eprint = "2112.05806",
    archivePrefix = "arXiv",
    primaryClass = "hep-ex",
    doi = "10.1140/epjc/s10052-022-10608-2",
    journal = "Eur. Phys. J. C",
    volume = "82",
    number = "8",
    pages = "694",
    year = "2022"
}

@Article{Ahlen1978,
     author    = "Ahlen, S.P.",
     title     = "{Stopping-power formula for magnetic monopoles}",
     journal   = "Phys. Rev. D",
     volume    = "17",
     pages     = "229",
     year      = "1978",
     doi       = "10.1103/PhysRevD.17.229", 
}

@Article{Ahlen1980,
     author    = "Ahlen, S.P.",
     title     = "{Theoretical and experimental aspects of the energy loss of relativistic heavily ionizing particles}",
     journal   = "Rev. Mod. Phys.",
     volume    = "52",
     pages     = "121",
     year      = "1980",
     doi       = "10.1103/RevModPhys.52.121",
}

@article{Song2022,
    author = "Song, Wen Yi and Taylor, Wendy",
    title = "{Pair production of magnetic monopoles and stable high-electric-charge objects in proton\textendash{}proton and heavy-ion collisions}",
    eprint = "2107.10789",
    archivePrefix = "arXiv",
    primaryClass = "hep-ph",
    doi = "10.1088/1361-6471/ac3dce",
    journal = "J. Phys. G",
    volume = "49",
    number = "4",
    pages = "045002",
    year = "2022"
}

@article{NNPDF23LO,
    author = "Carrazza, Stefano and Forte, Stefano and Rojo, Juan",
    title = "{Parton Distributions and Event Generators}",
    eprint = "1311.5887",
    archivePrefix = "arXiv",
    primaryClass = "hep-ph",
    doi = "10.48550/ARXIV.1311.5887",
    year = "2013"
}

@article{Manohar_2017,
    author = {Aneesh V. Manohar and Paolo Nason and Gavin P. Salam and Giulia Zanderighi},
    title = {The photon content of the proton},
    journal = {Journal of High Energy Physics},
    eprint = "1708.01256",
    archivePrefix = "arXiv",
    primaryClass = "hep-ph",
    doi = "10.1007/jhep12(2017)046",
    volume = "2017",
	number = "12",
	year = "2017"
  }

@Inproceedings{RooFitConf,
    author         = "Verkerke, Wouter and Kirkby, David",
    title          = "{The RooFit toolkit for data modeling}",
    year           = "2003",
    booktitle = "{Conference in Computing in High Energy and Nuclear Physics (CHEP03)}",
    doi = "10.5170/CERN-2011-006.313",
    eprint         = "physics/0306116",
    archivePrefix  = "arXiv",
    primaryClass   = "physics.data-an",
}

@Article{Run2lumi,
    author         = "{ATLAS Collaboration}",
    title          = "{Luminosity determination in \(pp\) collisions at \(\sqrt{s} = 13\,\text{TeV}\) using the ATLAS detector at the LHC}",
%    journal        = "Eur. Phys. J. C",
%    volume         = "76",
    year           = "2022",
%   pages          = "653",
    doi            = "10.48550/arXiv.2212.09379",
    eprint         = "2212.09379",
    archivePrefix  = "arXiv",
    primaryClass   = "hep-ex",
}

@article{Birks1952,
	Author = {Birks, J. D.},
	Journal = {Phys. Rev.},
	Pages = {569},
	Title = {{Theory of the Response of Organic Scintillation Crystals to Short-Range Particles}},
	Volume = {86},
	Year = {1952},
    doi = "10.1103/PhysRev.86.569",
 }

@article{HIPBirks,
	Author = {Burdin, S. and Horbatsch, M. and Taylor, W.},
	Doi = {10.1016/j.nima.2011.10.044},
	Journal = {Nucl. Inst. Meth. A},
	Pages = {111},
	Title = {{A correction to Birks' Law in liquid argon ionization chamber simulations for highly ionizing particles}},
	Volume = {664},
	Year = {2012}
 }

@article{Kurochkin2006,
    author = "Kurochkin, Yu. and Satsunkevich, I. and Shoukavy, Dz. and Rusakov\
ich, N. and Kulchitsky, Yu.",
    title = "{On production of magnetic monopoles via $\gamma \gamma$ fusion at\
 high energy $p p$ collisions}",
    doi = "10.1142/S0217732306022237",
    journal = "Mod. Phys. Lett. A",
    volume = "21",
    pages = "2873",
    year = "2006"
}

@ARTICLE{Dougall2009,
    author = "Dougall, T. and Wick, S. D.",
    title = "Dirac magnetic monopole production from photon fusion in proton collisions",
    journal = "Eur. Phys. J. A",
    year = "2009",
    volume = "39",
    pages = "213",
    doi = "10.1140/epja/i2008-10701-8",
    archivePrefix = "arXiv",
    eprint = "0706.1042",
    primaryClass = "hep-ph",
}

@article{Epele2012,
   author = "Epele, Luis N. and Fanchiotti, Huner and García Canal, Carlos A. and Mitsou, Vasiliki A. and Vento, Vicente",
   title = "Looking for magnetic monopoles at {LHC} with diphoton events",
   journal = "Euro. Phys. J. Plus",
   volume = "127",
   pages = "60",
   ISSN = "2190-5444",
   url = "http://dx.doi.org/10.1140/epjp/i2012-12060-8",
   doi = "10.1140/epjp/i2012-12060-8",
   number = "5",
   year = "2012",
   archivePrefix = "arXiv",
   eprint = "1205.6120",
   primaryClass = "hep-ph",
}

@article{Baines2018,
    author         = "Baines, S. and Mavromatos, N. E. and Mitsou, V. A. and Pinfold, J. L. and Santra, A",
    title          = "Monopole production via photon fusion and Drell-Yan processes: {MADGRAPH} implementation and perturbativity via velocity-dependent coupling and magnetic moment as novel features",
    journal        = "Eur. Phys. J. C",
    volume         = "78",
    year           = "2018",
    pages          = "966",
    doi            = "10.1140/epjc/s10052-018-6440-6",
    archivePrefix  = "arXiv",
    eprint         = "1808.08942",
    primaryClass   = "hep-ph",
    note = "[Erratum: Eur. Phys. J. C 79, 166 (2019)]",
}


%% file: ATLAS-useful.bib
@Article{Sjostrand:2006za,
     author    = "Sj{\"o}strand, Torbjorn and Mrenna, Stephen and Skands, Peter Z.",
     title     = "{PYTHIA 6.4 physics and manual}",
     journal   = "JHEP",
     volume    = "05",
     year      = "2006",
     pages     = "026",
     eprint    = "hep-ph/0603175",
     archivePrefix = "arXiv",
     doi       = "10.1088/1126-6708/2006/05/026",
     SLACcitation  = "%%CITATION = HEP-PH/0603175;%%"
}

@article{Sjostrand:2014zea,
      author         = "Sj{\"o}strand, Torbj{\"o}rn and Ask, Stefan and Christiansen,
                        Jesper R. and Corke, Richard and Desai, Nishita and Ilten,
                        Philip and Mrenna, Stephen and Prestel, Stefan and
                        Rasmussen, Christine O. and Skands, Peter Z.",
      title          = "{An introduction to PYTHIA 8.2}",
      journal        = "Comput. Phys. Commun.",
      volume         = "191",
      year           = "2015",
      pages          = "159",
      doi            = "10.1016/j.cpc.2015.01.024",
      eprint         = "1410.3012",
      archivePrefix  = "arXiv",
      primaryClass   = "hep-ph",
      reportNumber   = "LU-TP-14-36, MCNET-14-22, CERN-PH-TH-2014-190,
                        FERMILAB-PUB-14-316-CD, DESY-14-178, SLAC-PUB-16122,
                        --FERMILAB-PUB-14-316-CD",
      SLACcitation   = "%%CITATION = ARXIV:1410.3012;%%"
}

@Article{Alwall:2014hca,
      author         = "Alwall, J. and Frederix, R. and Frixione, S. and Hirschi,
                        V. and Maltoni, F. and Mattelaer, O. and Shao, H. -S. and
                        Stelzer, T. and Torrielli, P. and Zaro, M.",
      title          = "{The automated computation of tree-level and
                        next-to-leading order differential cross sections, and
                        their matching to parton shower simulations}",
      journal        = "JHEP",
      volume         = "07",
      year           = "2014",
      pages          = "079",
      doi            = "10.1007/JHEP07(2014)079",
      eprint         = "1405.0301",
      archivePrefix  = "arXiv",
      primaryClass   = "hep-ph",
      reportNumber   = "CERN-PH-TH-2014-064, CP3-14-18, LPN14-066, MCNET-14-09,
                        ZU-TH-14-14",
      SLACcitation   = "%%CITATION = ARXIV:1405.0301;%%"
}

@article{Agostinelli:2002hh,
      author         = "Agostinelli, S. and others",
      title          = "{\textsc{Geant4} -- a simulation toolkit}",
      journal        = "Nucl. Instrum. Meth. A",
      volume         = "506",
      year           = "2003",
      pages          = "250",
      doi            = "10.1016/S0168-9002(03)01368-8",
      reportNumber   = "SLAC-PUB-9350, FERMILAB-PUB-03-339",
      SLACcitation   = "%%CITATION = NUIMA,A506,250;%%"
}

@Article{Read:2002hq,
    Author = {Read, Alexander L.},
    Journal = {J. Phys. G},
    Pages = {2693},
    doi = "10.1088/0954-3899/28/10/313",
    Title = "{Presentation of search results: the \(CL_S\) technique}",
    Volume = {28},
    Year = {2002}
}


%% file: ATLAS.bib
@Article{PERF-2007-01,
    author         = "{ATLAS Collaboration}",
    title          = "{The ATLAS Experiment at the CERN Large Hadron Collider}",
    journal        = "JINST",
    volume         = "3",
    year           = "2008",
    pages          = "S08003",
    doi            = "10.1088/1748-0221/3/08/S08003",
    primaryClass   = "hep-ex",
}

@Article{EXOT-2011-27,
    author         = "{ATLAS Collaboration}",
    title          = "{Search for Magnetic Monopoles in \(\sqrt{s} = 7\,\text{TeV}\) \(pp\) Collisions with the ATLAS Detector}",
    journal        = "Phys. Rev. Lett.",
    volume         = "109",
    year           = "2012",
    pages          = "261803",
    doi            = "10.1103/PhysRevLett.109.261803",
    reportNumber   = "CERN-PH-EP-2012-177",
    eprint         = "1207.6411",
    archivePrefix  = "arXiv",
    primaryClass   = "hep-ex",
}

@Article{EXOT-2012-10,
    author         = "{ATLAS Collaboration}",
    title          = "{Search for long-lived, multi-charged particles in \(pp\) collisions at \(\sqrt{s} = 7\,\text{TeV}\) using the ATLAS detector}",
    journal        = "Phys. Lett. B",
    volume         = "722",
    year           = "2013",
    pages          = "305",
    doi            = "10.1016/j.physletb.2013.04.036",
    reportNumber   = "CERN-PH-EP-2012-360",
    eprint         = "1301.5272",
    archivePrefix  = "arXiv",
    primaryClass   = "hep-ex",
}

@Article{EXOT-2013-20,
    author         = "{ATLAS Collaboration}",
    title          = "{Search for heavy long-lived multi-charged particles in \(pp\) collisions at \(\sqrt{s}=\,8\,\text{TeV}\) using the ATLAS detector}",
    journal        = "Eur. Phys. J. C",
    volume         = "75",
    year           = "2015",
    pages          = "362",
    doi            = "10.1140/epjc/s10052-015-3534-2",
    reportNumber   = "CERN-PH-EP-2015-049",
    eprint         = "1504.04188",
    archivePrefix  = "arXiv",
    primaryClass   = "hep-ex",
}

@Article{EXOT-2014-16,
    author         = "{ATLAS Collaboration}",
    title          = "{Search for magnetic monopoles and stable particles with high electric charges in \(8\,\text{TeV}\) \(pp\) collisions with the ATLAS detector}",
    journal        = "Phys. Rev. D",
    volume         = "93",
    year           = "2016",
    pages          = "052009",
    doi            = "10.1103/PhysRevD.93.052009",
    reportNumber   = "CERN-PH-EP-2015-174",
    eprint         = "1509.08059",
    archivePrefix  = "arXiv",
    primaryClass   = "hep-ex",
}

@Article{PERF-2014-07,
    author         = "{ATLAS Collaboration}",
    title          = "{Topological cell clustering in the ATLAS calorimeters and its performance in LHC Run~1}",
    journal        = "Eur. Phys. J. C",
    volume         = "77",
    year           = "2017",
    pages          = "490",
    doi            = "10.1140/epjc/s10052-017-5004-5",
    reportNumber   = "CERN-PH-EP-2015-304",
    eprint         = "1603.02934",
    archivePrefix  = "arXiv",
    primaryClass   = "hep-ex",
}

@Article{IDET-2015-01,
    author         = "{ATLAS Collaboration}",
    title          = "{Performance of the ATLAS Transition Radiation Tracker in Run~1 of the LHC: tracker properties}",
    journal        = "JINST",
    volume         = "12",
    year           = "2017",
    pages          = "P05002",
    doi            = "10.1088/1748-0221/12/05/P05002",
    reportNumber   = "CERN-EP-2016-311",
    eprint         = "1702.06473",
    archivePrefix  = "arXiv",
    primaryClass   = "hep-ex",
}

@Article{PERF-2015-07,
    author         = "{ATLAS Collaboration}",
    title          = "{Study of the material of the ATLAS inner detector for Run~2 of the LHC}",
    journal        = "JINST",
    volume         = "12",
    year           = "2017",
    pages          = "P12009",
    doi            = "10.1088/1748-0221/12/12/P12009",
    reportNumber   = "CERN-EP-2017-081",
    eprint         = "1707.02826",
    archivePrefix  = "arXiv",
    primaryClass   = "hep-ex",
}

@Article{TRIG-2016-01,
    author         = "{ATLAS Collaboration}",
    title          = "{Performance of the ATLAS trigger system in 2015}",
    journal        = "Eur. Phys. J. C",
    volume         = "77",
    year           = "2017",
    pages          = "317",
    doi            = "10.1140/epjc/s10052-017-4852-3",
    reportNumber   = "CERN-EP-2016-241",
    eprint         = "1611.09661",
    archivePrefix  = "arXiv",
    primaryClass   = "hep-ex",
}

@Article{EXOT-2017-13,
    author         = "{ATLAS Collaboration}",
    title          = "{Search for heavy long-lived multicharged particles in proton--proton collisions at \(\sqrt{s} = 13\,\text{TeV}\) using the ATLAS detector}",
    journal        = "Phys. Rev. D",
    volume         = "99",
    year           = "2019",
    pages          = "052003",
    doi            = "10.1103/PhysRevD.99.052003",
    reportNumber   = "CERN-EP-2018-284",
    eprint         = "1812.03673",
    archivePrefix  = "arXiv",
    primaryClass   = "hep-ex",
}

@Article{EXOT-2017-20,
    author         = "{ATLAS Collaboration}",
    title          = "{Search for magnetic monopoles and stable high-electric-charge objects in \(13\,\text{TeV}\) proton--proton collisions with the ATLAS detector}",
    journal        = "Phys. Rev. Lett.",
    volume         = "124",
    year           = "2020",
    pages          = "031802",
    doi            = "10.1103/PhysRevLett.124.031802",
    reportNumber   = "CERN-EP-2019-084",
    eprint         = "1905.10130",
    archivePrefix  = "arXiv",
    primaryClass   = "hep-ex",
}

@Booklet{ATL-SOFT-PUB-2021-001,
    author         = "{ATLAS Collaboration}",
    title          = "{The ATLAS Collaboration Software and Firmware}",
    howpublished   = "{ATL-SOFT-PUB-2021-001}",
    url            = "https://cds.cern.ch/record/2767187",
    year           = "2021",
}

@Booklet{ATL-SOFT-PUB-2023-001,
    author         = "{ATLAS Collaboration}",
    title          = "{ATLAS Computing Acknowledgements}",
    howpublished   = "{ATL-SOFT-PUB-2023-001}",
    url            = "https://cds.cern.ch/record/2869272",
    year           = "2023",
}


%% file: CMS.bib
@Article{CMS-EXO-12-026,
    author         = "{CMS Collaboration}",
    title          = "{Searches for long-lived charged particles in \(pp\) collisions at \(\sqrt{s} = 7\) and \(8\,\text{TeV}\)}",
    journal        = "JHEP",
    volume         = "07",
    year           = "2013",
    pages          = "122",
    doi            = "10.1007/JHEP07(2013)122",
    reportNumber   = "CERN-PH-EP-2013-073",
    eprint         = "1305.0491",
    archivePrefix  = "arXiv",
    primaryClass   = "hep-ex",
}

@Article{CMS-EXO-15-010,
    author         = "{CMS Collaboration}",
    title          = "{Search for long-lived charged particles in proton--proton collisions at \(\sqrt{s} = 13\,\text{TeV}\)}",
    journal        = "Phys. Rev. D",
    volume         = "94",
    year           = "2016",
    pages          = "112004",
    doi            = "10.1103/PhysRevD.94.112004",
    reportNumber   = "CERN-EP-2016-204",
    eprint         = "1609.08382",
    archivePrefix  = "arXiv",
    primaryClass   = "hep-ex",
}


%% file: PubNotes.bib
@Booklet{ATL-PHYS-PUB-2014-021,
    author         = "{ATLAS Collaboration}",
    title          = "{ATLAS Pythia~8 tunes to \(7~\text{TeV}\) data}",
    howpublished   = "{ATL-PHYS-PUB-2014-021}",
    url            = "https://cds.cern.ch/record/1966419",
    year           = "2014",
}
